\documentstyle[12pt]{article}
\newcommand\mysection{\setcounter{equation}{0}\section}

\def\baeq{\begin{appeq}}     \def\eaeq{\end{appeq}}
\def\baeeq{\begin{appeeq}}   \def\eaeeq{\end{appeeq}}
\newenvironment{appeq}{\beq}{\eeq}
\newenvironment{appeeq}{\beeq}{\eeeq}
\def\bAPP#1#2{
 \markright{APPENDIX #1}
 \addcontentsline{toc}{section}{Appendix #1: #2}
 \medskip
 \medskip
 \begin{center}      {\bf\LARGE Appendix #1 :}{\quad\Large\bf #2} \end{center}
 \renewcommand{\thesection}{#1.\arabic{section}}        
\setcounter{equation}{0}
        \renewcommand{\thehran}{#1.\arabic{hran}}
\renewenvironment{appeq}
  {  \renewcommand{\theequation}{#1.\arabic{equation}}
     \beq
  }{\eeq}
\renewenvironment{appeeq}
  {  \renewcommand{\theequation}{#1.\arabic{equation}}
     \beeq
  }{\eeeq}
\nopagebreak \noindent}

\def\eAPP{\renewcommand{\thehran}{\thesection.\arabic{hran}}}

\renewcommand{\theequation}{\thesection.\arabic{equation}}
\newcounter{hran}
\renewcommand{\thehran}{\thesection.\arabic{hran}}

\def\bmini{\setcounter{hran}{\value{equation}}
\refstepcounter{hran}\setcounter{equation}{0}
\renewcommand{\theequation}{\thehran\alph{equation}}\begin{eqnarray}}
\def\bminiG#1{\setcounter{hran}{\value{equation}}
\refstepcounter{hran}\setcounter{equation}{-1}
\renewcommand{\theequation}{\thehran\alph{equation}}
\refstepcounter{equation}\label{#1}\begin{eqnarray}}

\def\emini{\end{eqnarray}\relax\setcounter{equation}{\value{hran}}\renewcommand{\theequation}{\thesection.\arabic{equation}}}

\def\cO#1{{\cal{O}}\left(#1\right)}
\def\pw#1{^{#1}}
\def\MSbar{\relax\ifmmode\overline{\rm MS}\else{$\overline{\rm MS}${ }}\fi}
\def\la{\mathrel{\mathpalette\fun <}}

\def\fun#1#2{\lower3.6pt\vbox{\baselineskip0pt\lineskip.9pt
  \ialign{$\mathsurround=0pt#1\hfil##\hfil$\crcr#2\crcr\sim\crcr}}}

\def\ie{\hbox{\it i.e.}{}}      
\def\eg{\hbox{\it e.g.}{}}      
\def\eV{{\rm e\kern-0.12em V}}            
 \def\GeV{{\rm G}\eV} 
\def\half{{\textstyle {1\over2}}}

\def \al {\relax\ifmmode{\alpha}\else{$\alpha${ }}\fi}
\def \be {\relax\ifmmode{\beta}\else{$\beta${ }}\fi}
\def\Im{\mathop{\rm Im}}    \def\Re{\mathop{\rm Re}}
\def\partder#1{{\partial   \over\partial #1}}
\def\abs#1{\left| #1\right|}
\def\lrang#1{\left\langle #1 \right\rangle}

\def\ben{\begin{enumerate}}  \def\een{\end{enumerate}}
\def\bit{\begin{itemize}}    \def\eit{\end{itemize}}
\def\beq{\begin{equation}}   \def\eeq{\end{equation}}
\def\beeq{\begin{eqnarray}}  \def\eeeq{\end{eqnarray}}
\def\bq{\begin{quote}}       \def\eq{\end{quote}}

\def\kp{\relax\ifmmode{k_\perp}\else{$k_\perp${ }}\fi}
\def\kps{\relax\ifmmode{k_\perp\pw2}\else{$k_\perp\pw2${ }}\fi}
\def \as{\relax\ifmmode\alpha_s\else{$\alpha_s${ }}\fi}
\def \pt{\relax\ifmmode{p_t}\else{$p_t${ }}\fi}

\def\eps{\relax\ifmmode\epsilon\else{$\epsilon${ }}\fi}

\def\ee{\relax\ifmmode{e\pw+e\pw-}\else{${e\pw+e\pw-}${ }}\fi}
\def\qq{\relax\ifmmode{q\overline{q}}\else{$q\overline{q}${ }}\fi}
\def\br{bremsstrahlung}
\newskip\humongous \humongous=0pt plus 1000pt minus 1000pt
\def\caja{\mathsurround=0pt}
\def\eqalign#1{\,\vcenter{\openup1\jot
\caja   \ialign{\strut \hfil$\displaystyle{##}$&$
\displaystyle{{}##}$\hfil\crcr#1\crcr}}\,}

\newif\ifdtup

\def\eqal2#1{\,\vcenter{\openup1\jot
\caja   \ialign{\strut \hfil$\displaystyle{##}$&\hfil$
\displaystyle{{}##}$\hfil &$
\displaystyle{{}##}$\hfil\crcr#1\crcr}}\,}

\def\np#1#2#3{{\em Nucl.Phys.}~\underline{B#1} (19#3) #2}
\def\pl#1#2#3{{\em Phys.Lett.}~\underline{#1B} (19#3) #2}
\def\pr#1#2#3{{\em Phys.Rev.}~\underline{D#1} (19#3) #2}
\def\prep#1#2#3{{\em Phys.Rep.}~\underline{#1} (19#3) #2}
\def\prl#1#2#3{{\em Phys.Rev.Lett.}~\underline{#1} (19#3) #2}

\def\spj#1#2#3{{\em Sov.Phys.JETP}\/~\underline{#1} (19#3) #2}

\def\zp#1#2#3{{\em Zeit.Phys.}~\underline{C#1} (19#3) #2}
\def\ijmp#1#2#3{{\em Int. J. Mod. Phys.}\/ {\underline {#1}} (19#3) #2}
 \def\cite#1{[\ref{#1}]}
 \def\citd#1#2{[\ref{#1},\ref{#2}]}
 \def\citt#1#2#3{[\ref{#1},\ref{#2},\ref{#3}]}
 \def\citq#1#2#3#4{[\ref{#1},\ref{#2},\ref{#3},\ref{#4}]}
 \def\citm#1#2{[\ref{#1}--\ref{#2}]}
\def\Apr{A^\prime}
\def\App{A^{\prime\prime}}

\def\cZ{{\cal{Z}}}
\def\cF{{\cal{F}}}
\def\cK{{\cal{K}}}

\def\cP{{\cal{P}}}
\def\cV{{\cal{V}}}
\def\cM{{\cal{M}}}
\def\cR{{\cal{R}}}
\def\at{\al_{\mbox{\scriptsize eff}}}

\def\aPT{\as^{\mbox{\scriptsize PT}}}
\def\FPT{F^{\mbox{\scriptsize PT}}}
\def\FNP{F^{\mbox{\scriptsize NP}}}
\def\NP{{\mbox{\scriptsize NP}}}
\def\tFPT{\tilde F^{\mbox{\scriptsize PT}}}

\def\L{\Lambda}
\def\PT{{\mbox{\scriptsize PT}}}
\def\Li{\mbox{Li}_2}
\def\cRr{{\cal{R}}^{(r)}}
\def\cFr{{\cal{F}}^{(r)}}
\def\re#1{(\ref{#1})}

\begin{document}
\begin{titlepage}
\begin{flushright}
CERN-TH/95-281 \\
Cavendish-HEP-95/12\\
hep-ph/9512336
\end{flushright}              
\vspace*{\fill}
\begin{center}
{\Large\bf Dispersive Approach to Power-Behaved\\[3mm]
Contributions in QCD Hard Processes\footnote{Research supported in 
part by
the UK Particle Physics and Astronomy Research Council and by the EC 
Programme
``Human Capital and Mobility", Network ``Physics at High Energy 
Colliders",
contract CHRX-CT93-0357 (DG 12 COMA).}}
\end{center}
\par \vskip 5mm
\begin{center}
        {\bf Yu.L.\ Dokshitzer\footnote{On leave from  
        St Petersburg Nuclear Physics Institute, Gatchina, 
        St Petersburg 188350, Russia.}}\\
        Theory Division, CERN, CH-1211 Geneva 23, Switzerland\\
        \vskip 0.3 cm
        {\bf G.\ Marchesini}\\
        Dipartimento di Fisica, Universit\`a di Milano and\\
        INFN, Sezione di Milano, Italy
        \vskip 0.1 cm
        and\\
        \vskip 0.1 cm
        {\bf B.R.\ Webber}\\
        Cavendish Laboratory, University of Cambridge, UK
        \end{center}
\par \vskip 2mm
\begin{center} {\large \bf Abstract} \end{center}
\begin{quote}
We consider power-behaved contributions to hard
processes in QCD arising from non-perturbative effects
at low scales which can be described by introducing
the notion of an infrared-finite effective coupling.
Our method is based on a dispersive treatment which
embodies running coupling effects in all orders.
The resulting power behaviour is consistent
with expectations based on the operator product
expansion, but our approach is more widely
applicable. The dispersively-generated power
contributions to different observables are
given by (log-)moment integrals of a universal
low-scale effective coupling, with
process-dependent powers and coefficients.
We analyse a wide variety of quark-dominated processes
and observables, and show how the power contributions
are specified in lowest order by the behaviour of one-loop
Feynman diagrams containing a gluon of small virtual
mass.  We discuss both collinear safe observables 
(such as the \ee total cross section and $\tau$ hadronic width,
DIS sum rules, \ee event shape variables and the Drell-Yan $K$-factor)
and collinear divergent quantities (such as DIS structure functions,
\ee fragmentation functions and the Drell-Yan cross section). 
\end{quote}
\vspace*{\fill}
CERN-TH/95-281\\
December 1995
\end{titlepage}

\mysection{Introduction}

Power-behaved contributions to hard collision observables are 
by now widely recognized both as a serious difficulty in improving 
the precision of tests of perturbative QCD and as a way to explore
non-perturbative effects. Such contributions are manifest as
power-varying discrepancies between fixed-order perturbative 
predictions and experiment.
For some quantities \citm{hadro}{AkZak}, 
the leading power contributions are of order
$\L/Q$ where $Q$ is the relevant energy scale and $\L\sim 300$
MeV is the characteristic scale of QCD. In this case,
even at scales $Q\sim M_Z\gg \L$, power-behaved terms can
be comparable to $\cO{\as^2}$ perturbative contributions.
In other processes 
\citm{CS}{NW}, such as deep inelastic lepton scattering
(DIS) and Drell-Yan lepton pair production, contributions of order
$\L^2/Q^2$ are important because $Q$ is not so large. At
even lower scales, as in the hadronic decay of the $\tau$
lepton, a stronger suppression of power contributions is
needed in order for a perturbative analysis to be
applicable \citm{BNP}{Neu}.

Such contributions are usually regarded as ``power-suppressed
corrections'' to the perturbative prediction, in which context
they might appear negligible in comparison with perturbative terms of
any order. From this viewpoint the former could not be considered 
before summing up the entire perturbative series, which in itself is
ambiguous. In practice, however, power contributions can easily be
distinguished since they are rapidly varying in contrast to the slowly
(logarithmically) varying fixed-order perturbative contributions.
Thus we can rather regard them as {\em power enhanced}, in the
sense that they grow more rapidly as the hard process scale $Q$
decreases. In view of their phenomenological relevance,
it is important to study any possible source of power-behaved
terms which might be a general feature of non-perturbative QCD.
 
So far, there is no rigorous systematic method for calculating
power contributions in QCD. In some cases the operator product
expansion (OPE) provides a framework for the classification of
possible power-behaved terms, but it gives little indication of their
coefficients \citd{OPE}{SVZ}.  General studies of the divergence of 
the perturbation series in high orders have led to the concept of
renormalons \cite{renormalons}, which are singularities in the
complex plane of the Borel variable conjugate to the running coupling.
In particular, infrared renormalons lie on the integration
contour of the Borel variable. In regularizing these singularities
one generates power-behaved contributions whose coefficients are 
ambiguous.
Formal mathematical manipulations alone cannot resolve 
the infrared renormalon problem, which is of a physical nature, namely 
the notion of the running coupling in the small $k^2$ regime.

In this context, it is natural to consider the possibility that 
a finite running coupling might be definable at low scales, at
least as an effective
measure of the strength of interaction at large distances.
One may hope in this way to take into account the main effects of
confinement at a sufficiently inclusive level.  This idea can only
be meaningful if the effective coupling so introduced is found to
be universal in some approximation. Its specific form, $\at(k^2)$,
at small $k^2$ then determines the power contributions.
Thus the experimental analysis of these contributions will
constrain the form of $\at(k^2)$ at small scales.
The theorists' task is
to analyse the power contributions which are generated
by the non-perturbative regime of the running coupling.
To this end, 
one should study perturbative Feynman diagrams for hard
process observables, in which the scale of the running coupling
is not restricted to the hard region, but also runs into the
non-perturbative domain.

A technique for studying power contributions within the language
of perturbative QCD is to introduce a small gluon mass 
\citq{hadro}{BB}{BBB}{BBZ}, which may be used as a
``trigger'' for long-distance contributions.
In the present paper, we attempt to put this approach on
a somewhat clearer theoretical basis. We work with a gluon
field of zero mass, and introduce a dispersive representation
of the running coupling $\as(k^2)$, which is assumed to
remain finite all the way down to $k^2=0$. 
In this way $\as(k^2)$ is expressed in terms of an effective coupling 
$\at(\mu^2)$ depending on the dispersion variable $\mu$.
In this representation $\mu$ plays a role similar to
a gluon mass as far as the phase space and Feynman denominators
are concerned. The region of small values of $\mu$
is responsible for power contributions. Since the
running coupling is by assumption universal, these contributions
can be related for different processes.
We study the corrections to a wide range of
quark-dominated processes due to a single off-shell gluon
with time-like or space-like virtuality.
The results should provide the means to test experimentally
the concept of a universal infrared-finite coupling
as an effective measure of strong interactions at large
distances, as reflected in sufficiently inclusive hard process
observables.
Our discussion is mainly in the context of Abelian gauge theory, 
although we argue that its application to the non-Abelian case 
may be justified for leading power contributions.  

In characterizing the effective coupling at low momentum scale 
we assume, in agreement with the standard ITEP-OPE approach 
\cite{SVZ},
that non-perturbative physics does not affect the high-momentum 
region, that is, the propagation of quarks and gluons at short distances
(soft confinement). 
Within our method, this implies that the non-perturbative 
``effective coupling modification'', $\delta\at(\mu^2)$, 
which is responsible for power corrections in hard distributions,
is essentially restricted to the small $\mu^2$ region 
and must not directly introduce power corrections into $\as$ itself
(notice again that $\mu^2$ is a dispersive variable and we are working 
in Minkowski space).  To satisfy this condition, 
$\delta\at(\mu^2)$ must have vanishing moments with respect 
to $\mu^2$, at least for the first few (integer) powers of $\mu^2$.
If this were not the case, the predictions of the operator
product expansion would be invalidated.

We start in Section 2 by discussing the extent to which a running
coupling might be definable beyond perturbation theory. 
We introduce a dispersive representation of $\as$ in terms of a spectral
density function $\rho_s(\mu^2)$ and then in terms of the effective 
coupling $\at(\mu^2)$.
We recall how this representation leads to a natural choice of
the scale of $\as$ in deep inelastic scattering, and we define
the corresponding `physical scheme' which we adopt for the
QCD coupling in the perturbative region.
Next we define the effective coupling $\at(\mu^2)$,
related to the spectral density function,
which will be a central quantity in our treatment
of power contributions.  

In Section 3 we present the method by which we compute
leading power contributions from integrals involving
the ``effective coupling modification'' 
$\delta\at$. We show that in quark-dominated processes, \
ie\ those involving no gluons at the Born level, 
the leading effect of the running
coupling is determined by the behaviour of $\at(\mu^2)$ and a
process-dependent {\em characteristic function} $\cF$. This
is a function of the ratio $\eps=\mu^2/Q^2$, together with
any relevant dimensionless ratios of hard scales $\{x\}$,
given by the sum of all one-loop graphs computed with a
non-zero gluon mass $\mu$.  The power contributions
that we seek to estimate are associated with the
non-analytic terms in the small-$\eps$ expansion
of this function. Each such term implies a power
correction proportional to an integral of
$\delta\at(\mu^2)$ times a corresponding
non-analytic weight function.  
Thus the task becomes to compute $\cF(\eps,\{x\})$ for
various processes and to see whether the predicted
power contributions are consistent with experiment
for some choice of the behaviour of the 
``effective coupling modification'' $\delta\at$ at low scales.

In Section 4 we present results on the characteristic
functions for a variety of processes and discuss
their behaviour at small $\eps$ and the associated
power contributions.  In many cases we find strong
enhancement factors, either logarithms of $Q^2$ or
singular functions of the auxiliary variables $\{x\}$,
which lead us to hope that these contributions may be the
dominant non-perturbative effects, at least in some
important regions of phase space. In the case of
$\L/Q$ contributions, no other possible sources of
such terms are known, and therefore our results
should provide useful semi-quantitative estimates
of their relative magnitudes in different observables.

We do not undertake any detailed
phenomenological analyses in the present paper,
but confine ourselves to providing the theoretical
results on which future analyses could be based.
We end with a summary of results and a discussion
of the limitations of our approach and its
possible further extensions and applications.

\mysection{Running coupling}
In hard processes initiated by a quark, the inclusive quantities at a 
given
order in perturbation theory receive contributions from an extra gluon
both from virtual corrections to the amplitude at the previous order
($k^2<0$) and from a new production channel ($k^2\ge 0$).
We discuss here how one can associate a running coupling with this gluon.
For large virtuality $\abs{k^2}$ it is clear how to
identify the running coupling and the proper scale for its argument
using perturbative methods.  For small momenta, $k^2 \to 0$,
the very language of quarks and gluon becomes scarcely applicable,
and the notion of the running QCD coupling becomes more elusive.
This situation is opposite to the case of the Abelian theory,
in which  the physical fine structure coupling is defined
at vanishing photon momentum, $\al(0)$, and appears, for
example, in the Thomson cross section.

We explore here the hypothesis that in a non-Abelian theory the
notion of the running coupling can also be extended to the region
of vanishing momenta, at least in some effective sense.
In doing so we seek to use the language and methods
of perturbative QCD to probe some non-perturbative phenomena, 
in particular the leading power contributions to hard cross sections.
In order to formulate this hypothesis more precisely,
we assume that in QCD the running coupling can be represented
by a dispersion relation which is inspired by the Abelian theory.
Therefore we first recall the case of QED and then describe the
extension to QCD.

\subsection{Space-like gluon}
\paragraph{Abelian case.}
Consider first the exchange of a photon with negative virtuality $-k^2$
(hereafter we always define $k^2>0$), which results in the appearance 
of the running coupling $\al(k^2)$.
In an Abelian theory, due to the simple form of the Ward identities
(\ie\ the cancellation of the fermion wave function and vertex
renormalization constants), the running coupling can be
reconstructed using dispersion techniques (see \cite{DS}).
The  dispersion relation for the running coupling reads
\beq\label{disrel}
\eqalign{
\al(k^2) \equiv \al(0) \cdot \cZ_3(-k^2)
&    = - \int_0^\infty \frac{d\mu^2}{\mu^2+k^2} \>\rho(\mu^2) 
\cr& = \al(0) +\!k^2 \int_0^\infty \frac{d\mu^2}{\mu^2}
\>\frac{\rho(\mu^2)}{\mu^2 +k^2} \>,}
\eeq
with $\cZ_3$ the photon wave function with the normalization 
$\cZ_3(0)=1$.
The spectral function $\rho(\mu^2)$ (positive in QED)
has support only on the positive real axis and
is obtained by considering
all discontinuities associated with the virtual photon.
It is given by the discontinuity of $\cZ_3$ at
positive virtuality,
\beq\label{spden}
\rho(\mu^2) \equiv  \frac{\al(0)}{2\pi i}
\left[\, \cZ_3(\mu^2-i\eps) - \cZ_3(\mu^2+i\eps)\, \right]
\>=\> -\frac{1}{2\pi i}\, \mbox{Disc}\> \left\{\al(-\mu^2)\right\}.
\eeq
The dispersive representation \re{disrel} has the following
non-perturbative implication:
the coupling $\al(k^2)$ is regular in the full $k^2$-complex plane
with a cut on the negative real axis.
Therefore this representation gives a coupling that is free from
the spurious QED Landau singularity 
and deviates from the standard perturbative e.m. coupling 
$\alpha^\PT(k^2)$ by a negligible amount $\cO{k^2/\Lambda^2}$ 
($\Lambda\sim 10^{30-40}\GeV$).

\paragraph{Non-Abelian case.}
Consider now the exchange of a gluon with negative virtuality
$-k^2$, which should result in the appearance of the running coupling
$\as(k^2)$.
In the QCD case one again seeks to identify the Feynman diagrams
responsible for the running coupling with the one-gluon reducible 
graphs,
which include gluon and quark self-energy insertions
together with vertex corrections. 
Such a class of diagrams does not constitute a gauge-invariant set.
Their gauge-dependent part is cancelled by corresponding
contributions from the one-gluon irreducible graphs (\eg, two-gluon 
exchange).
Therefore one can argue that these gauge-dependent
contributions from the one-gluon reducible set of graphs
remain finite at $k^2\!\to\!0$, while the gauge-invariant part
behaves as $1/k^2$, contributing to the renormalization of the pole
of the gluon propagator.
Since we are going to concentrate on the leading behaviour at small $k^2$,
we shall not analyse the contributions that are regular for $k^2 \to 0$.
We emphasize again that in non-Abelian
theories the very notion of one gluon exchange becomes elusive
at the level of relative $k^2$ corrections,
at which two gluons can mimic the one-gluon exchange,
the two processes being comparable and related by gauge invariance.

Inspired by the Abelian theory,
we shall assume that in QCD the running coupling $\as(k^2)$ still
satisfies a (formal) dispersion relation of the form \re{disrel}:
\beq\label{nonAb-disrel}
\as(k^2) \equiv \as(0) \cZ(-k^2)
= - \int\limits_0^\infty \frac{d\mu^2}{\mu^2+k^2} \>\rho_s(\mu^2)
\,, \quad   \rho_s(\mu^2) =
-\frac{1}{2\pi i}\, \mbox{Disc} \left\{\as(-\mu^2)\right\} .
\eeq
We note that the assumption of 
such a dispersion relation (with causal support)
implies the absence of the perturbative Landau pole at $k^2=\L^2$,
the QCD scale, 
but it also involves such ill-defined quantities as the coupling
$\as(0)$ and the spectral density $\rho_s$ at small scales $\mu^2$.
However, as we shall see shortly, $\as(0)$ will not appear explicitly
in physical observables, while the contribution of the spectral 
density in the small $\mu^2$ region will be suppressed by powers of $\mu^2$.

Since in QCD the cancellation between the vertex and fermion
propagator corrections due to Ward identities ($\cZ_\Gamma\cZ_q=1$)
does not hold, the quantity $\cZ$ in \re{nonAb-disrel}
is not simply the gluon propagator correction.
As is well known, both the non-Abelian part of the vertex 
renormalization
correction $\cZ^{\mbox{\scriptsize (na)}}_\Gamma$ and the gluon 
propagator
factor $\cZ_g$ contribute (in a gauge-dependent way) in forming the
running coupling. Therefore in the case of QCD
the quantity $\cZ$ in \re{nonAb-disrel} is now given by the product
\beq\label{Z}
\cZ \>=\> \cZ^{\mbox{\scriptsize (na)}}_\Gamma
\cdot \cZ_g \cdot \cZ^{\mbox{\scriptsize (na)}}_\Gamma\>.
\eeq
This expression shows that the function $\rho_s(\mu^2)$ is not
literally a spectral density (although we shall continue to call it
by that name) since it is not necessarily positive. In QED the
function $\rho(\mu^2)$ is given by the discontinuity of the
photon propagator, so that it is positive for any value of $\mu^2$ as
required by unitarity. This implies that $\alpha(k^2)$ increases as
$k^2$ increases.
In QCD instead $\as(k^2)$ decreases with increasing $k^2$ and so
$\rho_s(\mu^2)$ is negative, at least in the perturbative domain.
This is not in conflict
with unitarity since, according to \re{Z}, the function
$\rho_s(\mu^2)$ is given by the discontinuity of the gluon propagator
correction ($\cZ_g$) together with that of the non-Abelian vertex
corrections ($\cZ^{\mbox{\scriptsize (na)}}_\Gamma$).
For example, in a physical gauge (axial, planar; see \cite{DDT})
the first is determined by the total gluon splitting probability,
$$
\mbox{Disc}\,\cZ_g \propto \int_0^1 dz \left\{
2N_c\left(\frac1z+\frac{1}{1-z}-2+z(1-z)\right)
+\sum_1^{n_f}\left(z^2+(1-z)^2\right) \right\}
$$
and 
gives an infrared-divergent contribution, which is positive as
required by unitarity.
The second (vertex corrections) gives a (divergent) {\em negative}\/
contribution, 
$$
2\,\mbox{Disc}\,\cZ^{\mbox{\scriptsize (na)}}_\Gamma\>
\propto \> -4N_c\int_0^1 \frac{dz}{z}\>,
$$
so that the total discontinuity
is finite and gauge-independent but not positive definite:
$$\eqalign{
\mbox{Disc}\left( \cZ^{\mbox{\scriptsize (na)}}_\Gamma\cZ_g 
 \cZ^{\mbox{\scriptsize (na)}}_\Gamma\right) \>&\propto\>
\int_0^1 dz\left\{ 2N_c\left(-2+z(1-z)\right)
+\sum_1^{n_f}\left(z^2+(1-z)^2\right) \right\}\cr
&= -\frac{11}{3}N_c + \frac23n_f =-\beta_0\,.
}$$

\subsection{Time-like gluon}
We consider now the case of a gluon with zero ($k^2=0$) or positive 
virtuality ($k^2 >0$) corresponding to the contribution of new 
production channels. In inclusive quantities the integration over $k^2$
has to be performed up to some kinematic limit $k_{\max}^2\sim Q^2$, 
the relevant hard scale.
From this additional gluon correction one reconstructs the running 
coupling, provided one is able to factorize the following combination
\beq\label{as+}
\as(0)\,\delta(k^2) \>+\>
\frac{\rho_s(k^2)}{k^2}\>=\>
\frac{\as(0)}{\pi}\Im\left\{\frac{\cZ(k^2)}{-k^2-i\varepsilon}\right\}\,.
\eeq
The first contribution is due the on-shell gluon while the second 
is due to the sum over all final states generated by a
positive-virtuality gluon.
In order to factorize this combination, we need to consider 
observables which are sufficiently inclusive that all new production 
channels contribute to the same value of the observable.

Operationally our approach is similar to the so-called
``naive non-Abelianization'' procedure based on resummation
of fermion loop diagrams \citd{BBB}{Abel}.
The usual motivation for concentrating on fermion loop contributions
is to avoid problems with gauge invariance. However we believe that
the argument given earlier, that gauge-dependent terms are less
singular at $k^2\to 0$, justifies the application of our method
to the {\em leading}\/ power corrections in QCD.

In the next subsection we describe how the running coupling emerges in
inclusive quantities of this type. The dispersive
representation (\ref{as+}) allows one to
identify the proper scale for the argument of the running coupling,
and to define a `physical' scheme for the definition of the
QCD scale $\L$.

\subsection{Argument of the coupling}
For the sake of illustration we consider the non-singlet quark
distribution $F(Q^2,x)$ in DIS \cite{DS}.
Here $p$ and $q$ are the momenta of the incoming quark and of the
hard photon probe respectively, so that $Q^2=-q^2$ and $x=Q^2/2(pq)$ 
is the Bjorken variable. 

The evolution equation for $F$ is obtained by considering the emission 
of an additional gluon of momentum $k$. For simplicity we discuss only
the real emission contribution; virtual corrections can be included in
a straightforward way. 
One has to take into account both the on-shell contribution
at $k^2=0$, with coupling $\as(0)$, and the contribution from
the continuum involving the discontinuity through multi-parton
states, with strength given by the spectral density $\rho_s(k^2)$.
Introducing a minimum transverse momentum $Q_0$ for the emitted 
partons, the non-singlet quark distribution $F(Q^2,Q_0^2,x)$ is 
obtained by convoluting the combination \re{as+} with the
matrix element squared $\cF(Q^2,k^2,\kps,x,z)$  for emission
of an additional off-shell gluon of virtuality $k^2$, transverse 
momentum $\kps$ and fraction of momentum $(1-z)=2k\cdot 
(q+xp)\,x/Q^2$.
We have 
\beq\label{finres}
F(Q^2,Q_0^2,x)\>
=\int_x^1\frac{dz}{z} \> \int_{Q_0^2}^{Q^2}  {d\kps} \>
\int_{0}^{Q^2}  {dk^2} \> \cF(Q^2,k^2,\kps,x,z)\>
\left( \as(0)\delta(k^2)+\frac{\rho_s(k^2)}{k^2} \right),
\eeq
where the integrations over \kps and the timelike gluon virtuality $k^2$ 
actually run up to the kinematical limit, $W^2=Q^2(1-x)/x\sim Q^2$.

To see how the running coupling emerges in the {\em anomalous dimension}\/
it suffices to consider the quasi-collinear region $\kps\ll Q^2$.  
In the collinear limit the function $\cF$ can again be expressed 
in terms of the distribution $F$ itself as follows
\beq\label{cFApp}
\cF(Q^2,k^2,\kps,x,z)\> \simeq
F(Q^2,\kps,\frac x z)\>\frac{C_F}{2\pi} \left(
\frac{P(z)}{\kps+zk^2}-\frac{(1-z)\,zk^2}{(\kps+zk^2)^2}\right).
\eeq
The first term in the brackets, involving the standard $q\to q$ 
splitting function $P(z)=(1+z^2)/(1-z)$, is related to the $k^2=0$ 
limit of the scattering matrix element.
The second (subleading) contribution, non-singular for $k^2\to0$,
originates from the $k^2$-corrections to the matrix element.

In order to show how the integral over the virtual boson mass $k^2$
in (\ref{finres}) reproduces an effective coupling
running with \kps as scale, it suffices to perform two steps,
namely to extend the $k^2$ integration to infinity and then
to apply the dispersion relation.
The extension of the integration region is possible since the
$k^2$ integral is rapidly convergent above
$k^2 \sim \kps/z \la  \kps/x$ which, in the collinear approximation,
is much smaller than the kinematical boundary $k^2\le W^2\sim Q^2$.

Given this simplification, one can use the dispersion relation
(\ref{as+}) to obtain for the $k^2$ integrals in (\ref{finres})
\beq\label{argkt}\eqalign{
\int_{0}^{\infty}  {dk^2} \>\left(
\frac{P(z)}{\kps+zk^2}-\frac{(1-z)\,zk^2}{(\kps+zk^2)^2}\right)
&
\left(\as(0)\delta(k^2)+\frac{\rho_s(k^2)}{k^2}\right)
\cr 
& = P(z)\frac{\as(\kps/z)}{\kps}
-(1-z)\frac{d\as(\kps/z)}{d\kps}\,,
}\eeq
and the evolution equation becomes
\beq\label{impeveq}
F(Q^2,Q_0^2,x)
=\int_x^1\frac{dz}{z} \int_{Q_0^2}^{Q^2}  \frac{d\kps}{\kps} \>
F( Q^2,\kps,\frac xz) \frac{C_F}{2\pi} \left\{P(z) \,\as(\frac{\kps}{z})
- (1\!-\!z) \,\frac{d \as(\frac{\kps}{z}) }{d\ln\kps} \right\} .
\eeq
Thus, by using the spectral function and the dispersion relation
\re{as+}, we have reconstructed the running coupling associated with 
an extra gluon with positive virtuality.
Moreover, as in \citt{DDT}{GL72}{ABCMV},
we have shown that the dispersion relation allows one to identify the
physical scale for the argument of the coupling.

It is clear that in this way one has included some important
contributions from higher orders proportional to the first beta
function coefficient $\be_0$. In particular in \re{impeveq} one finds
the following contributions:
\begin{enumerate}
\item The last term in the curly bracket of \re{impeveq} can be
written as
$$
- (1-z) \,\frac{d \as(\kps/z) }{d\ln\kps}
=  (1-z) \,\frac{\be_0}{4\pi} \as^2(\kps)
+ \ldots \>,
\quad \left(\beta_0= \frac{11}{3}N_c -\frac23n_f\right)\; ;
$$

\item Expanding the argument of the coupling in the main term, we 
obtain
$$
  P(z)\,{\as(\kps/z)} = P(z)\,{\as(\kps)} + P(z)\, \ln z \;
\left(\frac{\be_0}{4\pi}\right){\as^2 (\kps)} \>+\> \ldots 
$$
\end{enumerate}
The above terms of order $\as^2$ are indeed present 
in the two-loop contribution to the anomalous dimension \cite{2loop}.

\paragraph{QCD scale.}
Two comments are in order concerning the ``physical scale'' 
of the coupling.

\par\noindent
1) Rewriting the expression in the curly brackets in (\ref{impeveq}) as
\beq
 P(z) \,\as\>  \>-\>  (1-z) \,\frac{d\, {\as}}{d\ln\kps}
\>=\> \left\{ \frac{2z}{1-z}\cdot \as
\>+\> (1-z)\cdot\left[\,\as + \frac{\be_0}{4\pi}\,\as^2\,\right] 
\>\right\},
\eeq
one observes that beyond the first loop the coupling of the two pieces
of the gluon radiation probability $P(z)=2z/(1-z)+(1-z)$ effectively
acquire different arguments. This is because these two contributions 
to the splitting function are physically different.
The first term comes from the universal ``soft'' \br\
distribution ($d\omega/\omega$), which is independent
of the nature of the incoming parton and corresponds to
the gluon polarization in the scattering plane (longitudinal polarization),
while the second one is due to ``hard'' gluons ($\omega\,d\omega$),
depends on initial parton spin and consists equally of
longitudinal and transverse polarization.

\par\noindent
2) The universal nature of {\em soft}\/ gluon \br\ may be used to 
define a ``physical'' QCD coupling beyond the first loop.
To do so one has to analyse the higher-order terms of the anomalous 
dimension and to absorb the singular contributions $\propto (1-z)^{-1}$ 
systematically into a redefinition of the QCD scale $\Lambda$, 
which determines the intensity of radiation of relatively soft gluons.

Such a physical scheme is related to the popular \MSbar scheme as follows.
In the \MSbar scheme the intensity of soft \br\ is given by an infinite 
series in a formally defined parameter $\al_{\MSbar}$. In particular,
both the first and the second loop \MSbar splitting
functions $P(z)$ and $P^{(2)}_{\MSbar}(z)$ contain the
``soft singularity'' $(1-z)^{-1}$.
Redefining the coupling in such a way that the new two-loop
splitting function $P^{(2)}(z)$ does not contain the soft
singularity (the so-called MC-scheme of \cite{CMW})
one arrives, in second order, at the relation
\bminiG{phcointr}
\label{twosch}
 \al_{\MSbar}P(x) + \al^2_{\MSbar}\,P^{(2)}_{\MSbar}(x)
& =& \as\, P(x) + \as^2\, P^{(2)}(x)   \>+\>\cO{\as^3}\>; \\
\label{phco}
\as  = \al_{\MSbar}\left( 1+ \frac{\al_{\MSbar}}{2\pi}\,\cK \right)\>,
\quad &&
\quad \Lambda = \Lambda_{\MSbar} \exp(\cK/\be_0)\;,
\eeeq
with
\beeq
\label{cKdef}
 {\cK} &=& C_A\left( \frac{67}{18} -\frac{\pi^2}{6}\right)
 -\frac{10}{9} n_fT_R  \;,
\emini
which is equal to $4.0\pm 0.5$ for $n_f=4\mp 1$.

In this scheme the second loop contribution $P^{(2)}(z)$
is at least one power of $(1\!-\!z)$ down with respect to the
singular part of the main \br\ spectrum, $P(z)$.
Moreover, as advocated also in Ref.~\cite{BLM}, 
$P^{(2)}(z)$ no longer contains any $n_f$-dependence,
this ill-defined quantity being completely absorbed, to this order, 
into the momentum dependence of the coupling (for more details see 
\cite{DKT}).

Because of the classical character of soft \br,
such a definition of the coupling proves to be universal with respect 
to the nature of the source of radiation. In particular the two-loop 
{\em gluon}\/
splitting function also becomes infrared regular in the physical 
scheme described above, \ie,
$P^{(2)}_{g\to gg}(z)/P_{g\to gg}(z)\to 0$ with $z\!\to\!1$.
For this reason we shall use this physical scheme for the QCD coupling.

\subsection{Effective coupling}
The spectral density $\rho_s$ may be related to the running
coupling by the following formal transformation\footnote{Let us recall that
we are considering here the coupling constant $\as(k^2)$ which is free from 
spurious singularities in the Euclidean domain, $k^2>0$.}:
\beq\label{trick}\eqalign{
\as(k^2)
&= -\int_0^\infty \frac{d\mu^2}{\mu^2+k^2}\>
\rho_s(\mu^2) = -\int_0^\infty \frac{dv}{1+v}\>\rho_s(vk^2)
\cr&
= -\int_0^\infty \frac{dv}{1+v}\>\exp\left\{
\ln v\frac{d}{d\ln k^2} \right\}\cdot \rho_s(k^2)
\cr&
= \frac{\pi}{\sin\left(\pi \frac{d}{d\ln k^2}\right)}\> \rho_s(k^2)\>.
}\eeq
In terms of the differential operator $\cP$,
\beq
\cP \equiv \frac{d}{d\ln \mu^2}\>,
\eeq
one may thus write the inverse relation as
\beq
 \rho_s(\mu^2) = \frac{1}{\pi}\sin(\pi\cP)\, \as(\mu^2)\>.
\eeq
Introducing the {\em effective coupling} $\at(\mu^2)$ defined by
\beq\label{aeffrel}
 \rho_s(\mu^2) =\frac{d}{d\ln\mu^2}\> \at(\mu^2) \>,
\eeq
we have
\beq\label{atint}
\as(k^2) = k^2 \int_0^\infty \frac{d\mu^2}{(\mu^2+k^2)^2}\>\at(\mu^2)
\eeq
with inverse
\beq\label{atdef}
 \at(\mu^2) \equiv  \frac{\sin(\pi\cP)}{\pi\cP} \> \as(\mu^2)\>.
\eeq

It follows from Eq.~\re{atdef} that in the perturbative domain $\as\ll 
1$,
the standard and effective couplings are approximately the same:
\beq\label{atexpn}
 \at(\mu^2) = \as(\mu^2) -\frac{\pi^2}{6}\frac{d^2\as}{d\ln^2\mu^2}
 + \ldots \>=\> \as\,+\, \cO{\as^3}\>.
\eeq
In what follows we shall look upon $\at$ itself as an effective 
measure of QCD
interaction strength which extends the physical perturbative coupling
down to the non-perturbative domain.

\mysection{Dispersive method}

Consider a dimensionless inclusive quantity $F(Q^2,\{x\})$
for a hard process involving only quarks at the Born level.
Here $Q^2$ is the hard scale and $\{x\}$ stands for any further 
relevant
dimensionless parameters, \eg, Bjorken $x$ or its conjugate
moment variable $N$ (DIS structure functions),
particle energy fraction (inclusive \ee annihilation spectra),
the ratio $M^2/s$ (the Drell-Yan process), or a jet shape variable.
The one-loop prediction for $F(Q^2,\{x\})$ , obtained from the
squared amplitudes involving one additional gluon,
can be improved by using a dispersive method based on the
representation of the running coupling in \re{nonAb-disrel} for 
the space-like case and \re{as+} for the time-like case.
In this way one takes into account relevant higher-order perturbative
effects and is also able to study power contributions arising from
the non-perturbative behaviour of the coupling at low scales.

\subsection{Basic equation}
To describe the method we concentrate first on collinear safe 
quantities.
The generalization to include collinear singular observables will be
presented shortly. We have to consider the two cases in which the 
additional
gluon contributes to real production channels and to a virtual 
correction.

Consider first the case in which the additional gluon contributes to 
new production channels. As discussed in Sect.\ 2.2, one has to take 
into account both the on-shell contribution at $k^2=0$, with coupling 
$\as(0)$, and the contribution from the continuum involving the 
discontinuity through multi-parton states (from gluon branching), with
strength given by  the spectral density $\rho_s(k^2)$.
For all these contributions at fixed $k^2$ one would like to factorize 
the combination \re{as+}. This is possible only if the observable
is fully inclusive with respect to gluon branching,
as in the case of total cross sections, DIS structure functions, etc.
For less inclusive quantities, such as jet shape
observables, gluon branching may give a different contribution
\cite{NS}. The non-factorizable contribution is higher-order in $\at$.
However, since $\at$ enters at a low scale, such terms could still
be comparable to the factorizable first-order contribution 
$\propto\at$.
Therefore our method can only give quantitative predictions
for less inclusive quantities if $\at$ at low scales is sufficiently 
small.

We denote by $\cF^{\mbox{\scriptsize real}}(Q^2,\{x\},k^2)$ the 
squared amplitude for the emission of a gluon with timelike virtuality 
$k^2 \ge 0$, apart from the coupling.
$\cF^{\mbox{\scriptsize real}}(Q^2,\{x\},k^2)$ is defined to
vanish outside the real phase space $0<k^2<k_{\max}^2(Q^2,\{x\})$.
Then the real emission contribution to the inclusive quantity
$F(Q^2,\{x\})$ is given by the sum of the on-shell and continuum 
contributions (see Eq.~\re{as+}):
\beq\label{realem}
F^{\mbox{\scriptsize real}}(Q^2,\{x\})
=\as(0)\cF^{\mbox{\scriptsize real}}(0)
+ \int_0^\infty \frac{dk^2}{k^2}\rho_s(k^2)
\cdot\cF^{\mbox{\scriptsize real}}(k^2)\,.
\eeq
Consider now the case in which the additional gluon contributes to 
virtual corrections.
This contribution is present when the observable under
consideration does not vanish in the Born approximation.
In this contribution the virtuality of the gluon is spacelike, 
$-k^2<0$,
and the coupling is given by $\as(k^2)$. The contribution can
then be written as
$$
F^{\mbox{\scriptsize virt}}(Q^2,\{x\})
=
\int_0^\infty
\frac {dk^2}{k^2}{\as(k^2)}\cM (Q^2,\{x\},-k^2)
\,,
$$
with $\cM$ the integrand for the Feynman diagram containing the
(massless) virtual gluon, apart from the coupling.
We now use the dispersive representation \re{nonAb-disrel} for
the running coupling to write
\beq\label{vir}
\eqalign{
F^{\mbox{\scriptsize virt}}
&
=
{\as(0)} \int_0^\infty
\frac {dk^2}{k^2}\cM(-k^2)
+\int_0^\infty\frac{d\mu^2}{\mu^2}\rho_s(\mu^2)
\int_0^\infty
\frac{dk^2}{k^2+\mu^2}\cM(-k^2)
\cr&
=\as(0)\cF^{\mbox{\scriptsize virt}}(0)
+ \int_0^\infty \frac{d\mu^2}{\mu^2}\rho_s(\mu^2)
\cdot\cF^{\mbox{\scriptsize virt}}(\mu^2)
\,,
}
\eeq
where
\beq
\cF^{\mbox{\scriptsize virt}}(Q^2,\{x\},\mu^2) \equiv
\int_0^\infty
\frac{dk^2}{k^2+\mu^2}\cM(Q^2,\{x\},-k^2)
\eeq
is the one-loop virtual correction that would be produced by a gluon
with a finite mass $\mu^2$.

The result thus has the same structure both for the real \re{realem}
and for the virtual \re{vir} contributions.
Combining them to form
$$
\cF(Q^2,\{x\},\mu^2) \equiv
\cF^{\mbox{\scriptsize real}}(Q^2,\{x\},\mu^2)
+\cF^{\mbox{\scriptsize virt}}(Q^2,\{x\},\mu^2)
\,,
$$
which we shall refer to as the {\it characteristic function},
the improved one-loop formula reads
(omitting the external variables $Q^2$ and $\{x\}$)
\beq\label{WDM1}
 F \>=\> \as(0)\cF(0) + \int_0^\infty 
\frac{d\mu^2}{\mu^2}\rho_s(\mu^2)
\cdot\cF(\mu^2)
\>=\> \int_0^\infty \frac{d\mu^2}{\mu^2}\rho_s(\mu^2)
\cdot
\left[\,\cF(\mu^2) - \cF(0)\,\right],
\eeq
where we have made use of the formal relation \re{nonAb-disrel} 
to eliminate $ \as(0)$.
By introducing the effective coupling $\at(\mu^2)$  
\re{atdef}
using Eq.~\re{aeffrel} and integrating by parts, we can write
\beq\label{WDM}
  F(Q^2,\{x\})\>=\>  \int_0^\infty
\frac{d\mu^2}{\mu^2}\> \at(\mu^2) \cdot \dot\cF (Q^2,\{x\};\mu^2) \>,
\;\;\;\;\;\;\;\;
\quad  \dot{\cF} \equiv -\frac{\partial\cF}{\partial\ln \mu^2}
\>.
\eeq
The relation \re{WDM1} or \re{WDM} is our basic equation for studying 
both the perturbative part and non-perturbative contribution to $F$.
For a collinear safe observable $\dot{\cF}(\mu^2)$ vanishes
for $\mu^2\to0$ and $\mu^2\to\infty$,
and the integral is dominated by the region $\mu^2\sim Q^2$.
One observes that the integrand contains two well separated scales: 
$Q^2$, the scale of the characteristic function $\cF(\mu^2)$, and
$\Lambda^2$, the QCD parameter, which is the scale of the effective 
coupling $\at(\mu^2)$ or $\rho_s(\mu^2)$. 
The perturbative contribution is related to the scale $Q^2$, while the 
power-behaved contribution is related to $\L^2$. 

Let us stress that although in our derivation we have introduced
the quantity $\as(0)$, which may make little sense in the context
of QCD, the final results \re{WDM1} or \re{WDM} involve only
the spectral density  $\rho_s(\mu^2)$ or the effective coupling
$\at(\mu^2)$ convoluted with a function that vanishes at $\mu^2=0$.
Therefore the basic equation may make sense even if $\as(0)$
does not exist.

In what follows we shall look upon $\at$ defined in \re{atdef} as an 
effective measure of QCD interaction strength which extends the 
physical perturbative coupling down to the non-perturbative domain.
While the form of $\at(k^2)$ at large $k^2$ is well known, we treat 
$\at$ at small scales as a phenomenological function whose 
behaviour is to be determined from experiment.

The characteristic function $\cF$ is process-dependent and is obtained
in practice by computing the relevant one-loop graphs with a
non-zero gluon mass $\mu$.  Note that we do {\em not} intend to
imply that the gluon has in any sense a real effective mass, but
only that the dispersive representation can be expressed in this way.  

In the above derivation it was implied that the
characteristic function is well-defined at
$\mu^2=0$.  This is true for collinear safe quantities such as
the total \ee annihilation cross section, DIS sum rules, the
Drell-Yan $K$ factor, jet shape variables, etc.
On the other hand for collinear singular quantities,
such as DIS structure functions, the Drell-Yan cross section,
and inclusive \ee parton fragmentation functions, we have
\beq\label{Freg}
 \cF(Q^2,\{x\};\mu^2) = P(\{x\})\cdot\ln\frac{Q^2}{\mu^2} \>+\>
  \cF^{\mbox{\scriptsize reg}}(Q^2,\{x\};\mu^2)\,,
\eeq
where $\cF^{\mbox{\scriptsize reg}}(Q^2,\{x\};0)$ is finite.
We cannot use the representations \re{WDM1} for the collinear
singular distribution itself, but rather for the scaling violation
rate $Q^2\partial F/\partial Q^2$. In this case the function
$\dot\cF$ is involved, which has a finite $\mu^2\to 0$ limit. Then
Eq.~\re{WDM} becomes
\beq\label{scaviol}
Q^2\frac{\partial}{\partial Q^2} F(Q^2,\{x\})
\>=\>  \int_0^\infty\frac{d\mu^2}{\mu^2}\>
\at(\mu^2) \cdot \ddot\cF (Q^2,\{x\};\mu^2) \>,
\eeq
where, from \re{Freg} we have
\beq
\ddot\cF (Q^2,\{x\};\mu^2)
\>=\>\left(\eps\frac{\partial}{\partial\eps}\right)^2
\cF^{\mbox{\scriptsize reg}}(\eps,\{x\})\>,
\eeq
and we have used the fact that $\dot\cF$ is a function of
$\eps=\mu^2/Q^2$.  The function $\ddot\cF$ vanishes as $\eps\to 0$
and $\eps\to\infty$, and therefore the integral in \re{scaviol}
is now well defined. For simplicity, we continue the
discussion in terms of $\cF$ and $\dot\cF$, with the understanding
that one should substitute $\dot\cF$ and $\ddot\cF$ when studying
the scaling violation in a collinear singular quantity.

We now discuss the general behaviour of the characteristic function.
Since $\cF$ depends only on dimensionless ratios, we write
\beq
\cF (Q^2,\{x\};\mu^2) =\cF (\eps,\{x\})\;,
\;\;\;\;\;\;
\dot{\cF} \equiv -\eps\frac{\partial}{\partial\eps}\cF (\eps,\{x\})\;,
\;\;\;\;\;\;
\eps\equiv\frac{\mu^2}{Q^2}\;.
\eeq
We describe the form of the characteristic function $\cF(\eps,\{x\})$
for a collinear safe quantity, and in particular its behaviour for
large and small $\eps$.
For a collinear singular quantity, the same behaviour is obtained for 
its regular part $\cF^{\mbox{\scriptsize reg}}(\eps,\{x\})$.
 
\paragraph{\boldmath Large $\eps$.}
At large $\eps$ the characteristic function either vanishes 
identically
because of the phase space boundary (in the absence of virtual
corrections to this order, as is the case for jet shape variables),
or decreases as a negative power of $\eps$ as a
consequence of the renormalizability of the theory.
It follows that the logarithmic derivative $\dot\cF$
also vanishes at infinity, so that the integral in Eq.~\re{WDM}
is well-defined.

\paragraph{\boldmath Small $\eps$.}
The behaviour at small $\eps$ is crucial for the analysis of
non-perturbative power contributions.
By definition of a collinear safe quantity, the
behaviour of $\dot\cF$ near $\eps=0$ is of the form
\beq\label{Fsmalleps}
\dot\cF(\eps) \>=\> \eps^{p}\,\left[ f(\ln\eps) + \eps\, g(\ln\eps)
+\cO{\eps^2} \right] \,,
\eeq
with $p>0$, where $f$ and $g$ are polynomials of degree
not higher than 2:
\beq\label{flneps}
f(\ln\eps) = f_2\ln^2\eps + f_1\ln\eps + f_0\;,
\eeq
and similarly for $g(\ln\eps)$.

As an example, we show in Figs.~\ref{Reefig1} and \ref{Reefig2}
the behaviour of $\cR\equiv 2\pi\cF/C_F$ and its derivative $\dot\cR$
for the \ee total annihilation cross section.  In this case we
have $\dot\cR\sim 3\eps^2+2\eps^3\ln\eps -3\eps^3$ at small $\eps$,
as shown by the dashed curve in Fig.~\ref{Reefig2},
so $p=2$, $f(\ln\eps)=3$ and $g(\ln\eps)=2\ln\eps-3$.

We show finally how the standard one-loop result is recovered from
Eq.~(\ref{WDM}).
Since $\dot\cF$ vanishes both for $\eps\to0$ and for $\eps\to\infty$,
the main contribution to the integral comes from $\eps\sim 1$,
that is $\mu^2\sim Q^2$.
Therefore one can extract the leading contribution by taking
the value of the effective coupling at some characteristic scale
$$
\bar{Q}^2\equiv c(\{x\})\cdot Q^2
$$
outside the integration, to obtain
\beq\label{F1int}
 F(Q^2,\{x\}) = \at(\bar{Q}^2)\cdot \cF(Q^2,\{x\};0) \>+\>
  \int_0^\infty
 \frac{d\mu^2}{\mu^2} \left[\, \at(\mu^2)-\at(\bar{Q}^2)\,\right]
 \cdot \dot\cF (Q^2,\{x\};\mu^2)\,.
\eeq
The remaining integral contributes both to higher-order perturbative 
terms and to non-perturbative power contributions.  
Roughly speaking, the former come from the region 
$\mu^2\sim\bar Q^2\sim Q^2$ and the latter from $\mu^2\sim \L^2\ll 
Q^2$.  
The small-$\mu^2$ region is outside
perturbative control but, to the extent that $\at$ may be defined
universally and obtained from experimental data, its contribution
is determined by the behaviour of $\cF$ at small values of $\eps$.
In the remainder of this Section, we consider
various approaches to the evaluation of the integral in \re{F1int}.

\subsection{Renormalons}
For large $Q^2$ the main contribution to the integral in \re{F1int} 
still comes from the region $\mu^2 \sim Q^2$.
One might take this as sufficient 
motivation for evaluating this contribution by expanding $\at(\mu^2)$ 
around $\mu^2=Q^2$ (for simplicity we consider here $\bar{Q^2}=Q^2$).
In this region the effective coupling $\at(\mu^2)$ can be approximated 
by its one-loop (or two-loop) perturbative expression $\aPT(\mu^2)$
\beq\label{aspert}
\at(\mu^2) \simeq \frac{4\pi}{\be_0\ln(\mu^2/\L^2)}
\equiv \aPT(\mu^2)
\,,
\;\;\;\;\;\;\;
\mu^2 \gg \L^2
\,.
\eeq
Making the replacement 
\beq\label{atexp}
\at(\mu^2)\Rightarrow 
\aPT(\mu^2)\,=\,\aPT(Q^2) \,+\,
\aPT(Q^2) \sum_{k=1}^\infty\, \left( \frac{\be_0\aPT(Q^2)}{4\pi}
\;\ln\frac{Q^2}{\mu^2}\right)^k
\,,
\eeq
one finds 
\beq\label{Cexp}
\eqalign{
F(Q^2)-\at({Q}^2)\cF(0) \Rightarrow 
&\aPT(Q^2) \sum_{k=1}^\infty 
\left(\frac{\be_0\aPT(Q^2)}{4\pi}\right)^{k}
\;C_k
\,,
\cr
C_k = 
&\int_0^\infty \frac{d\eps}{\eps} \left(\ln\frac{1}{\eps}\right)^k
\>\dot\cF(\eps)
\,,
}
\eeq
where the simplified notation $\cF(Q^2,\{x\},\mu^2)=\cF(\eps)$ has 
been used.
The deficiency of the perturbative expansion 
becomes apparent
from the fact that the resulting coefficients $C_k$ exhibit factorial 
growth \cite{renormalons}.
This is associated with both the ultraviolet and the infrared 
integration regions in $\eps$.

The ultraviolet contribution to $C_k$ is estimated by integrating 
\re{Cexp} over $\eps > 1$. If $\dot \cF(\eps)$ vanishes like
$\eps^{-q}$ at large $\eps$, one finds for large $k$
\beq
 C_k^{\mbox{\scriptsize UV}}
 \sim \int_1^\infty \frac{d\eps}{\eps} 
\left(\ln\frac{1}{\eps}\right)^k
\;\eps^{-q} \> \sim\>(-1)^k\,k!
\;.
\eeq
This corresponds to an ultraviolet renormalon.
Such an alternating series can be evaluated by Borel summation.
This is because in the ultraviolet integration region of (\ref{F1int}) 
the replacement $\at(\mu^2) \to \aPT(\mu^2)$ is a reliable 
approximation 
and the contribution of this region can in fact be evaluated 
explicitly 
without any expansion.

The infrared contribution to $C_k$ is estimated by integrating 
over the region $\eps<1 $. 
Using the small $\eps$ behaviour in \re{Fsmalleps} one finds 
\beq
C_k^{\mbox{\scriptsize IR}}
= \int_0^1 \frac{d\eps}{\eps} \left(\ln\frac{1}{\eps}\right)^k
\;\eps^{p}\,f(\ln \eps) \>\sim\>k!
\eeq
One again finds a factorial behaviour (an infrared renormalon). In 
this 
case however the coefficients are {\em non-alternating}\/ and therefore
the series is not Borel-summable.
Attempts to ascribe meaning to such a series by brute force tend to 
give
rise to unphysical {\em complex}\/ contributions at the level of 
$Q^{-2p}$ 
terms.  This is generally interpreted as an intrinsic uncertainty 
in the summation of the perturbative 
series\footnote{In fact, infrared renormalons are a purely perturbative
phenomenon and have no direct relation to the presence of the Landau
singularity in the running coupling; for a detailed discussion see
\cite{DU}.}.
It is important to recognize that formal mathematical manipulations alone
cannot resolve this problem, which is of a physical nature.
One requires genuinely new physical input, namely a power-behaved
`confinement' contribution, to obtain a sensible answer. 
In this paper we advocate the hypothesis that such an input, for
sufficiently inclusive Minkowskian observables, may be embodied in 
the form of $\at(\mu^2)$ at small $\mu^2$.

\subsection{Soft confinement and power corrections}
The standard operator product expansion (OPE)
approach by Shifman, Vainshtein and Zakharov \cite{SVZ}
quantifies confinement effects in terms
of additive contributions to Euclidean quantities 
and supplies the fuel for the impressive machinery of
the ITEP sum rules (for a review see \cite{ITEP}).
  
The OPE approach is based on the hypothesis that the
entire effect of confinement
in vacuum correlators of currents
may be embodied into ``condensates''
(vacuum expectation values of gauge- and Lorentz-invariant
colourless operators built from gluon and quark fields).
The basic ITEP idea was to separate the long- and short-distance
contributions and to treat them on different bases. 
In the Euclidean region such a separation
is straightforward: for $k^2=k_0^2+ |\vec{k}|^2>\lambda^2\sim 1\;\GeV^2$
one employs perturbation theory --- in particular,
one uses the purely perturbative expression $\aPT$ for the
coupling in Feynman diagrams for a given correlator.  The
complementary region $k^2<\lambda^2$, on the other hand,
is treated phenomenologically. Upon integration, the latter
region gives a power-behaved contribution of the order of
$(\lambda^2/Q^2)^p$.  

As far as gluon propagation is concerned, within the logic of the 
present
paper one may write equivalently
\beq\label{coupsplit}
  \as(k^2) \>=\> \aPT(k^2) + \delta\as(k^2)\,,
\eeq
with $\delta\as$ a modification in the effective interaction strength
at small momentum scales responsible for non-perturbative effects.
In agreement with the ITEP point of view, while this contribution 
generates power corrections to the hard distributions, it should not
modify the behaviour of the running coupling at large momentum scales.

Let us stress that the separation \re{coupsplit} does not mean that
the {\em perturbative}\/ contribution as such is free from 
power-behaved terms. This very question is almost meaningless:
one would have to keep under control an infinite series of logarithmic
high-order terms prior to addressing the problem of
perturbatively-generated
power corrections. As we have mentioned above,
due to the ``infrared renomalon problem'' the perturbative
series is apparently not summable to such a level of accuracy. 
At the same time, the separation prescription \re{coupsplit}
cures the problem operationally: as long as perturbative
integrals are cut off in the infrared, resummation of the
infinite series triggered by the running of $\aPT$ is harmless.    
To obtain a physically sensible (and reasonably accurate) answer 
it suffices to calculate a few terms of the perturbative expansion 
and then to add power-behaved terms as a contribution
of essentially different (non-perturbative, confinement) origin.  

There is a strong point of the OPE ideology,
which one may refer to as the ``{\em soft}\/ confinement'' scenario.
The OPE prescription implies that the propagation of quarks and gluons
with {\em large}\/ (Euclidean) momenta remains unaffected by 
non-perturbative
physics, even at the level of power-suppressed terms. 
Formally, one might argue that the propagators of coloured
fields have no gauge-invariant meaning. 
Nonetheless a power-suppressed variation, say a
$1+\lambda^2/k^2$ correction to the gluon propagator
in the {\em ultraviolet}\/ region $k^2\gg\lambda^2$, 
would inevitably introduce an additional ``{\em non-perturbative}'' 
contribution. 
This would have no relation to the region of small momentum 
flow in the corresponding Feynman diagrams for the current 
correlators,
which region is responsible for the formation of condensates.
As a result, the original separation idea would not be valid.
Thus the term $\delta\as(k^2)$ in \re{coupsplit} should decay
faster than some (sufficiently large negative) power 
$(k^2)^{-p_{\max}}$ for
viability of the notion of a condensate of dimension $2p\le 
2p_{\max}$.

Physically the OPE prescription corresponds to a picture of {\em smooth}\/ 
non-perturbative large-scale vacuum fields with a typical size
$\sim\lambda^{-1}$.
If such fields were non-singular at short distances, the gluon propagator
(and thus the running coupling) would be subject to exponentially 
small corrections only.
An instanton--anti-instanton gas as a representative model for 
non-trivial vacuum fields sets an upper bound 
$p_{\max}< \beta_0\sim 9$,  above which small-size instantons 
start to disturb the propagation of quarks and gluons.   

Following the discussion in the previous sections, we are now 
in a position to implement a similar logic for quantifying confinement
contributions to Minkowskian observables. 

The effective coupling $\at(\mu^2)$ appearing in \re{WDM1} and 
\re{WDM}
cannot be defined perturbatively below $\mu^2\sim \lambda^2\sim 1\;\GeV^2$,
where the very language of quarks and gluons is scarcely applicable.
In this region we expect an ``effective coupling modification'' 
$\delta\at$ which generates the non-perturbative interaction strength
$\delta\as$ in \re{coupsplit}
via the dispersion relation \re{atint}, \ie
\beq\label{alal}
\delta\as(k^2)= k^2\int_0^\infty \frac{d\mu^2}{(\mu^2+k^2)^2} \,
\delta\at(\mu^2)\>.
\eeq
At first sight, one might feel free to choose an arbitrary form for
$\delta\at$ at small scales to model confinement effects. 
However, on inspecting \re{alal} one observes that, generally 
speaking, 
a finite modification of the effective coupling at low scales will 
affect
the ultraviolet behaviour of the coupling in the Euclidean region 
by an amount proportional to $1/k^2$.

As discussed above, such a modification would ruin the basis of the 
OPE approach. 
One has therefore to require that at least the first $p_{\max}$ 
integer moments of this coupling vanish. A similar constraint has been
discussed in \cite{DSU}.
 
Consider, for example, the non-perturbative gluon condensate 
which contributes to the Adler $D$-function
(see, e.g., \cite{renormalons}).
To first order in $\as$ it is given by the integral
\beq\label{Gcond}
\frac{2\pi^2}{3}\lrang{\frac{\as}{\pi} G^2}_{\NP}
= \frac{3C_F}{2\pi}\int_0^{U^2} dk^2\,k^2\>\delta\as(k^2)\,,
\eeq
where $U^2$ is the ultraviolet cutoff.
Substituting the representation \re{alal} for $\delta\as$ 
in terms of the function $\delta\at$ and performing the
integration over $k^2$, one obtains
$$
\frac{2\pi^2}{3}\lrang{\frac{\as}{\pi} G^2}_{\NP} 
= \frac{3C_F}{2\pi}\int_0^\infty d\mu^2\, \delta\at(\mu^2)
\left[\, U^2 -2\mu^2\ln\frac{U^2}{\mu^2} + \mu^2\,\right].
$$
Convergence of the integral in \re{Gcond} implies
\beq\label{vanish}
 \int_0^\infty \frac{d\mu^2}{\mu^2}
\>\left(\mu^2\right)^p \delta\at(\mu^2)\>=\>0\>; \quad p=1,2\,, 
\eeq
that is, the vanishing of the first two moments of $\delta\at(\mu^2)$.
The result now reads
\beq\label{GGatrel}
\frac{\pi^2}{9}\, \lrang{\frac{\as}{\pi} G^2}_{\NP}
\>=\> \frac{C_F}{2\pi}\int_0^\infty
\frac{d\mu^2}{\mu^2}\,\mu^4\ln\mu^2\> \delta\at(\mu^2)\>.
\eeq
Notice this (log-)moment integral is independent of the scale 
of the logarithm. 

Inspecting higher moments (operators of higher dimension),
one finds it necessary to impose on $\delta\at(\mu^2)$ the set
of restrictions \re{vanish} with $p=1,2,\ldots p_{\max}$,
to respect the criterion of ``soft confinement'' as seen
through Euclidean eyes. Applied to a Minkowskian observable $F$, 
this means that only those terms in the small-$\mu^2$ behaviour
of the characteristic function $\dot\cF(\mu^2)$ 
that are {\em non-analytic}\/ in $\mu^2$ will contribute to $\FNP$.
Among them are the terms with non-integer $p$ and/or those with 
$\log\mu^2$-enhanced asymptotic behaviour.

Eq.\re{GGatrel} gives an example of a ``measurement'' of
one of the log-moments of $\delta\at$.
In the rest of the present paper we study how
this and other non-zero moments of $\delta\at$
enter into different Minkowskian observables.
According to Eq.~\re{WDM}, the corresponding contribution to
a generic collinear safe observable $F$ will be of the form
\beq
\FNP(Q^2,\{x\}) = \int_0^\infty \frac{d\mu^2}{\mu^2}\> 
\delta\at(\mu^2)\>
\dot\cF(Q^2,\{x\}; \mu^2)\,. 
\eeq
The leading non-zero contribution to the integral will come from
the first non-analytic term in the small-$\eps$ expansion of
$\cF (Q^2,\{x\};\mu^2) =\cF (\eps,\{x\})$, Eq.~\re{Fsmalleps}.
If $p$ is not an integer, the first term is non-analytic and
the leading contribution is proportional to $Q^{-2p}\ln^q Q$
where $f_q\ln^q\eps$
is the largest non-vanishing term in Eq.~\re{flneps}. If $p$ is
an integer and $f_1$ or $f_2$ is non-vanishing, the leading  
contribution is proportional to $Q^{-2p}\ln^{q-1} Q$, because the 
first singular contribution in $\mu^2$ will be of the form
$\mu^{2p}\ln \mu$. If $p$ is an integer and both
$f_1$ and $f_2$ are zero, the contribution will be proportional
to $Q^{-2(p+1)}\ln^{q-1} Q$, where $q\ge1$ is now specified by the
largest non-vanishing log-enhanced term of $g(\eps)$.

We shall call power-behaved contributions of this
type {\em dispersively-generated} power corrections.
An important question to be addressed is: how much of
the full power correction to a given observable
is due to these dispersively-generated terms?

For quantities like DIS structure functions and the Drell-Yan 
$K$-factor,
dispersively-generated $1/Q^2$ power contributions arise which are
given by the first (log- and log$^2$-enhanced) moment of 
$\delta\at(\mu^2)$.
The origin of these contributions may be traced back to a
{\em universal}\/ $1/Q^2$-suppressed non-perturbative correction
to the relevant {\em hard cross sections}\/ (coefficient functions).
At the same time, from general OPE considerations one expects
$1/Q^2$ corrections proportional to the hadronic matrix elements
of the relevant twist-4 operators, which depend on the target hadron.
Since the dispersively-generated terms have a definite, calculable
dependence on the hard process kinematics, it may be possible to
identify kinematic regions in which they are dominant.

In the case of \ee-annihilation, which is a
hard process free from initial-state hadrons,
the situation might be simpler.
In particular, for the practically important case of 
{\em linear}\/ ($1/Q$) corrections to event shapes,
dispersively-generated terms could describe the entire
contributions to different quantities in a universal way.
Such a hypothesis seems plausible not only because there is no competition
from the conventional OPE contributions, but also because it
follows naturally from the picture of soft (local in phase space) 
hadronization, which one invokes to explain the observed similarity 
(duality) between the calculable distributions of partons 
and the measured distributions of hadrons from QCD jets \cite{LPHD}.  

\mysection{Applications}

In this section we apply the dispersive method of the previous
section to various hard processes dominated by quarks.
Recall that the object of central importance is the
characteristic function $\cF(\eps)$ for the emission of a
gluon with mass-squared $\mu^2 = \eps Q^2$ at the hard scale $Q^2$.
For power corrections, the relevant contribution is given by the
leading non-analytic term in the small-$\eps$ behaviour of the
logarithmic derivative $\dot\cF$, which is of the
general form \re{Fsmalleps}. The coefficient of the
resulting contribution is then determined by the
function $f(\ln\eps)$ (or by the next-to-leading
function $g(\ln\eps)$ if the leading term is analytic).
Since $f$ or $g$ is a polynomial in $\ln(\mu^2/Q^2)$ of
degree not more than two, we introduce for future use the
moment integrals
\beq\label{adefs}
A_{2p} \>=\>\frac{C_F}{2\pi}
\int_0^\infty\frac{d\mu^2}{\mu^2}
\>\mu^{2p}\>\,\delta\at(\mu^2)
\eeq
and their derivatives
\beq\label{apdef}
\Apr_{2p} = \frac{d}{dp} A_{2p}\;,\;\;\;\;
\App_{2p} = \frac{d^2}{dp^2} A_{2p}\;,
\eeq
which have respectively an extra factor of $\ln\mu^2$ and $\ln^2\mu^2$ 
in the integrand of Eq.~\re{adefs}. Then all dispersively-generated
power corrections proportional to $Q^{-2p}$ can be represented 
as linear combinations of $A_{2p}$, $\Apr_{2p}$ and $\App_{2p}$. 
Since the integrand must be non-analytic in $\mu^2$, the unprimed moments
$A_{2p}$ can only contribute when $p\le p_{\max}$ 
is not an integer, while the primed (log) moments can contribute for any $p$. 
The quantity on the right-hand side of Eq.\re{GGatrel}, for example,
is denoted by $\Apr_4$, corresponding to a $1/Q^4$ correction to
the Adler $D$-function. 

In general the characteristic function $\cF(\eps)$ has both real and
virtual contributions. Since the latter are universal we discuss them first.
Their form depends on whether the momentum transfer in the hard 
process is space-like or time-like.

Consider the one-gluon virtual correction to a hard process (\eg\ DIS) 
with a space-like momentum transfer $-q^2=Q^2>0$. The total
correction to the renormalized hard interaction vertex is
of the form $C_F\as\cV_s/2\pi$ where
\bminiG{virts}\eqalign{
 \cV_s(\eps) &= -2\int_0^1dz(1-z)^2\int_0^\infty\frac{dk^2\>
Q^2}{(k^2+\mu^2)(k^2+zQ^2)}
= -2\int_0^1dz\frac{(1-z)^2}{z-\eps}\ln\frac{z}{\eps}  \cr
&= 2(1-\eps)^2\left[\Li(1-1/\eps) -\frac{\pi^2}{6} \right]  
-\frac72 -(3-2\eps)\ln\eps+2\eps \cr
&= 2(1-\eps)^2\left[\Li(\eps)+\ln\eps\ln(1-\eps)
-\frac 12\ln^2\eps -\frac{\pi^2}{3} \right]  
-\frac72 -(3-2\eps)\ln\eps+2\eps \,, }
\eeeq
where 
$$
\Li(u)\equiv -\int_0^u\frac{dt}{t}\ln(1-t) \,.
$$
For a time-like process  ($q^2=Q^2>0$, as in \ee annihilation and
the Drell-Yan process), the virtual correction is $C_F\as\cV_t/2\pi$ where
\beeq\label{virtt}\eqalign{
  \cV_t(\eps)\>&=\> \Re\, \cV_s(-\eps)
=\> -2\int_0^1 dz\,\frac{(1-z)^2}{z+\eps}\,\ln\frac{z}{\eps} \cr
&= 2(1+\eps)^2\left[\Li(-\eps)+\ln\eps\ln(1+\eps)
-\frac 12\ln^2\eps +\frac{\pi^2}{6} \right]  
-\frac72 -(3+2\eps)\ln\eps-2\eps \,.
}\emini
Next we combine these with the real contributions for 
various hard processes and analyse the resulting power corrections.
We consider first collinear safe and then collinear singular
processes.

\subsection{\boldmath Total \ee annihilation cross section}
\begin{figure}
\vspace{9.0cm}
\includegraphics{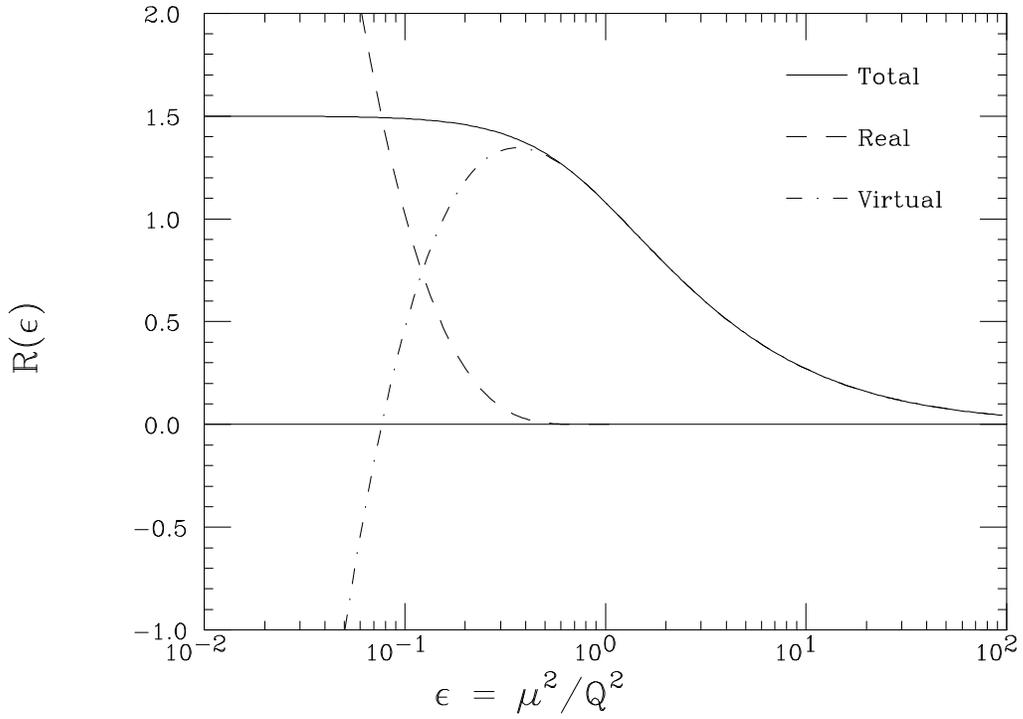}
\caption{Characteristic function for \ee total cross section, $\cR$.}
\label{Reefig1}
\end{figure}

The most straightforward application of the method is
to calculate the power correction to the \ee annihilation
cross section $R(Q^2)$, given to first order by
(we suppress the standard parton model normalization factor
$R^{0}=N_c\sum_f e_f^2$)
\beq\label{RPT}
  R^{(1)} = 1 + \frac{3C_F}{4\pi}\,{\as(Q^2)} + \ldots \;.
\eeq
In our approach, this result originates from perturbative
evaluation of the expression
\beq\label{RWDM}
  R\>=\> 1 + \frac{C_F}{2\pi} \int_0^\infty
 \frac{d\mu^2}{\mu^2} \at(\mu^2) \cdot 
\dot\cR (Q^2,\mu^2)
\,,\qquad
\dot{\cR}\equiv -\frac{d}{d\ln \mu^2}\cR(Q^2,\mu^2)\>
\,, 
\eeq
where $\cR$ is the characteristic function obtained from diagrams
involving one gluon with mass $\mu$. This function has both
a real and a virtual contribution (see Fig.~\ref{Reefig1}). 
The latter is simply given by the function $\cV_t(\eps)$ in 
\re{virtt}.
To obtain the real contribution we consider the emission of a quark,
an antiquark and a gluon of momenta $p$, $\bar p$ and $k$ respectively 
($k^2=\mu^2$, $p^2=\bar p^2=0$). The matrix element assumes a simple 
form in terms of the scaled quark and antiquark energies,
$$
x = 2pQ/Q^2\;,\;\;\;\; \bar x = 2\bar pQ/Q^2\;,
$$
which satisfy the phase-space constraints
$$
(1-x)(1-\bar x)\ge \eps\;,\;\;\;\; x+\bar x \le 1-\eps\;.
$$
The matrix element squared is then of the form $C_F\as \cM_{ee}/2\pi$, 
where
\beq\label{eeM}
\cM_{ee}(x,\bar x,\eps) =\frac{(x+\eps)^2+ (\bar x+\eps)^2 }
{(1-x)(1-\bar x)} -\frac{\eps}{(1-x)^2} -\frac{\eps}{(1-\bar x)^2} \,.
\eeq
The real contribution to the characteristic function $\cR$ is thus 
\beq\label{Rr}
\eqalign{
&
\cRr(\eps)=\int_{ph.sp.}d x\, d \bar x \cM_{ee}( x,\bar x, \eps) \cr
&
=-2(1+\eps)^2\left[\,2\Li(-\eps)+2\ln\eps\ln(1+\eps)-\frac 12\ln^2\eps
+\frac{\pi^2}{6}\,\right] 
+5 +(3+4\eps+3\eps^2)\ln\eps -5\eps^2
\,,}
\eeq
with $\eps \le 1$. At the edge of phase space 
($\eps \to  1$) the distribution vanishes rapidly, as 
\beq
 \cRr(\eps)= \frac1{10}(1-\eps)^5 \left\{1 +\cO{1-\eps}\right\}
\, .
\eeq
Finally the complete characteristic function is 
\beq\label{eeR}
\eqalign{
&
\mbox{for}\;\;\; \eps > 1 \,, \;\;\;\;
\cR(\eps)=\cV_t(\eps)\,, 
\cr&
\mbox{for}\;\;\; \eps < 1 \,, \;\;\;\;
\cR(\eps)=\cRr(\eps) +\cV_t(\eps)
\cr&
\;\;\;\;\;\;\;\;
=-2(1+\eps)^2\left[\,\Li(-\eps)+\ln\eps\ln(1+\eps) \,\right]
+\frac32 +(2+3\eps)\eps\ln\eps-2\eps-5\eps^2
\,.
}
\eeq

\begin{figure}
\vspace{9.0cm}
\includegraphics{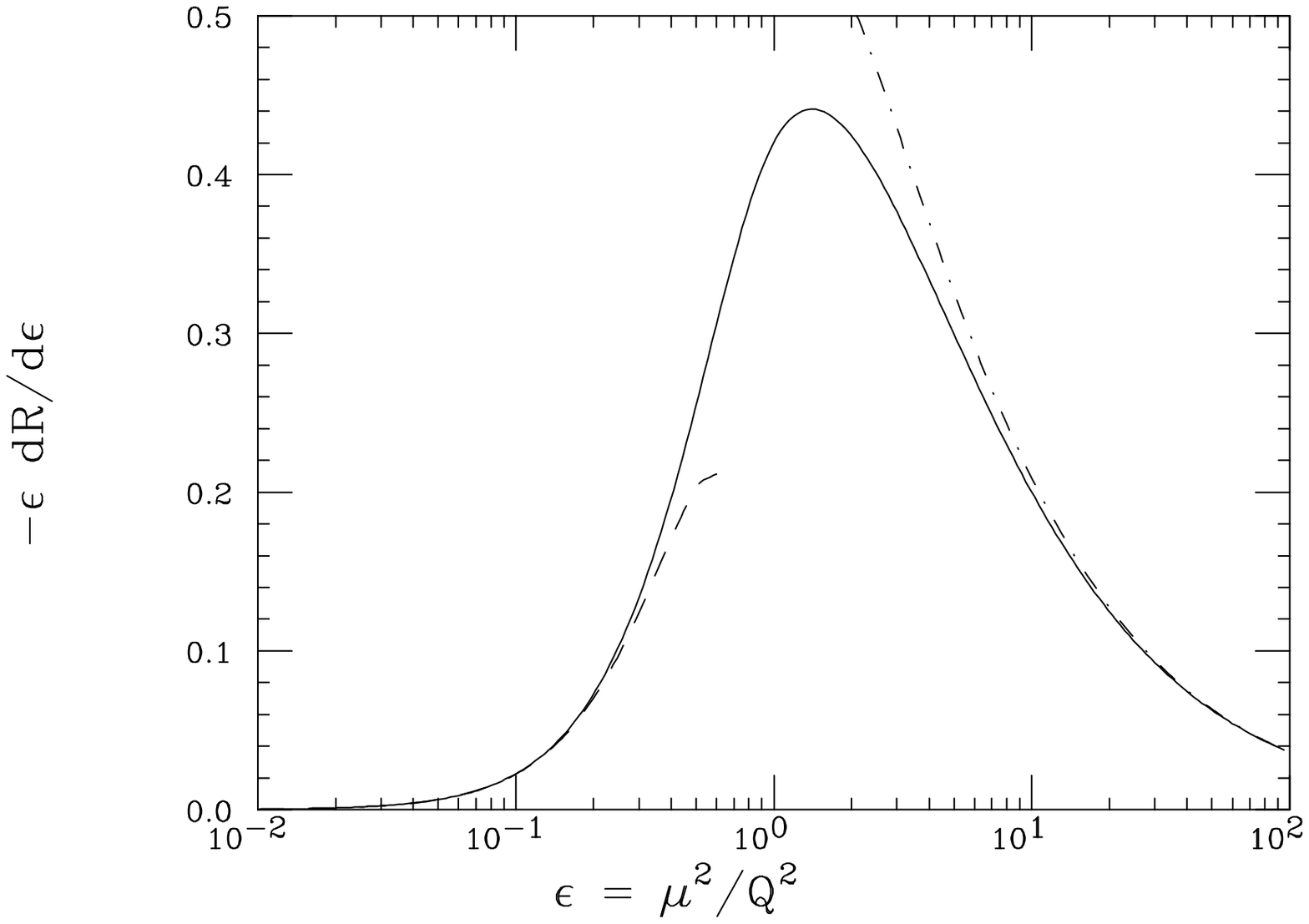}
\caption{Derivative of characteristic function for \ee total cross 
section,
$\dot\cR$. Dashed and dot-dashed curves show the limiting behaviour
at small and large $\eps$, respectively.}
\label{Reefig2}
\end{figure}

The function $\cR(\eps)$ is plotted in Fig.~\ref{Reefig1}
together with its logarithmic derivative $\dot\cR$ in 
Fig.~\ref{Reefig2}.
The behaviour for small and large $\eps$ is as follows:
\par\noindent
For $\eps\ll1$
\bminiG{as}
 \cR= \frac32(1-\eps^2)
-\frac{2\eps^3}3\left(\ln\eps-\frac{11}6\right)
\>+\>\cO{\eps^4\ln\eps}\, .
\eeeq
For $\eps\to\infty$ 
\beeq\label{leas}
 \cR= \frac2{3\eps}\left(\ln\eps+\frac{11}{6}\right) 
\>+\> \cO{\frac{\ln\eps}{\eps^2}}\,.
\emini
These two limiting forms are in fact related by
the following symmetry of the \ee characteristic function.

\paragraph{Symmetry.}
The function $\cR$ given in \re{eeR} satisfies 
the following inversion symmetry
\beq\label{SYMM}
\frac 1{\eps}\left[\cR(\eps)-\frac32\right]\>=\>
\eps\left[\cR\left(\frac1\eps\right)-\frac32\right]
\,.
\eeq
The behaviour for $\eps \to 0$
follows from this symmetry and the fact that
for $\eps\to\infty$ the characteristic function 
$\cR(\eps)=\cV_t(\eps)$ vanishes in the way specified in \re{leas}.

The symmetry \re{SYMM} also explains the fact that in the sum of real
and virtual contributions one finds that not only the terms singular
as $\mu^2 \to 0$,
$$
\ln^2\mu^2\;,  \;\;\;\;\;\; \ln\mu^2
\,,
$$
cancel, as required by the Bloch-Nordsieck theorem, but that the 
cancellation extends \citd{BB}{BBZ} also to the following finite terms
$$
\mu^2\ln^2\mu\;, \;\;\;\; \mu^2\ln\mu\,, \;\;\;\; \mu^2\;, 
\;\;\;\; \mu^4\ln^2\mu\;, \;\;\;\; \mu^4\ln\mu 
\,. 
$$

\paragraph{\boldmath Perturbative evaluation of $R(Q^2)$.}
Now we are in a position to evaluate $R$ according to (\ref{RWDM}).
Since $\dot\cR$ vanishes both for small and large $\eps$, 
\beq\label{cRas}
\dot\cR = 3\eps^2 +2\eps^3\ln\eps +\ldots \,, 
\;\;\;\;\eps\to0\;,
\;\;\;\;\;\;\;\;\;\;\;
\dot\cR=-\frac{2}{3\eps}{\ln\eps}+\ldots
\;\;\;\;\eps\to\infty
\,,
\eeq
the main contribution to \re{eeR} comes from $\eps\sim 1$, that is 
$\mu^2\sim Q^2$ (see Fig.~\ref{Reefig2}). 
Therefore the leading perturbative contribution to 
$R^{\mbox{\scriptsize PT}}(Q^2)$ is obtained by taking the value of 
the coupling constant $\at(\bar{Q}^2)$ outside the integration and one 
finds 
\beq
R^{\mbox{\scriptsize PT}}(Q^2)\,=\,1+\frac{C_F}{2\pi}
\at(\bar{Q}^2)\cR(0) \>+\> \cO{\as^2(\bar{Q}^2)} \>,
\eeq
where $\cR(0)=3/2$ and $\bar{Q}^2 \sim Q^2$ should be chosen in the 
vicinity of the peak, so that $\at(\bar{Q}^2) \approx\as({Q}^2)$
and one obtains the result \re{RPT}.

\paragraph{\boldmath Leading power correction.}
As indicated in Eq.~\re{cRas}, the first non-analytic term in the
expansion of $\dot\cR$ at small $\eps$ is of order $\eps^3\ln\eps$.
Thus the leading power-behaved contribution is given in terms of the
moment integral \re{adefs} for the effective coupling as
\beq\label{Rpc1}
R^{\mbox{\scriptsize NP}} \simeq 2 \frac{\Apr_6}{Q^6}\,.
\eeq
The absence of a $1/Q^2$ contribution follows
from the lack of any suitable dimension-two operators, in the
massless quark limit. In principle one might have expected that
a $1/Q^4$ contribution could be present, due to the gluon
condensate $\lrang{\as G^2}$. As we have already remarked,
such a contribution is indeed expected in the Adler $D$-function,
but to first order in $\delta\at$
it does not appear in $R$ itself, which is related to
the discontinuity of the $D$-function \cite{renormalons}.
In order for a $1/Q^4$
term to appear in $R$ one would need a $\ln Q^2/Q^4$ term in $D$.
At the same time, a leading power correction of the form \re{Rpc1}
is consistent with the OPE, since the $D$-function does acquire
a log-enhanced contribution from operators of dimension six 
(see \citd{BBB}{Neu}).

\paragraph{\boldmath Hadronic width of the $\tau$ lepton.}
The $\tau$ decay width \citm{BNP}{Neu} is closely related to
the \ee annihilation cross section. It is normally expressed
in terms of the quantity $R_{\tau}$, which for massless quarks is
\beq
R_{\tau} = 2\int_0^{m_{\tau}^2}\frac{ds}{m_{\tau}^2}\,
\left(1 -\frac{s}{m_{\tau}^2}\right)^2
\left(1+2\frac{s}{m_{\tau}^2}\right)R(s)\;.
\eeq
Thus we can write
\beq\label{Rtauint}
  R_{\tau}\>=\> 1 + \frac{C_F}{2\pi} \int_0^\infty
 \frac{d\mu^2}{\mu^2} \at(\mu^2) \cdot 
\dot\cR_{\tau} (m_{\tau}^2,\mu^2)
\eeq
where
\beq
\dot\cR_{\tau}(m_{\tau}^2,\mu^2)=2
\int_0^{m_{\tau}^2}\frac{ds}{m_{\tau}^2}
\left(1 -\frac{s}{m_{\tau}^2}\right)^2
\left(1+2\frac{s}{m_{\tau}^2}\right)
\dot\cR(s,\mu^2)\;.
\eeq
Defining $y=\mu^2/m_{\tau}^2$ and $\eps=\mu^2/s$, we can
write this in the form
\beq
\dot\cR_{\tau}(y)= 2y
\int_y^\infty\frac{d\eps}{\eps^2}
\left(1-\frac{y}{\eps}\right)^2
\left(1+2\frac{y}{\eps}\right)\dot\cR(\eps)\;.
\eeq

\begin{figure}
\vspace{9.0cm}
\includegraphics{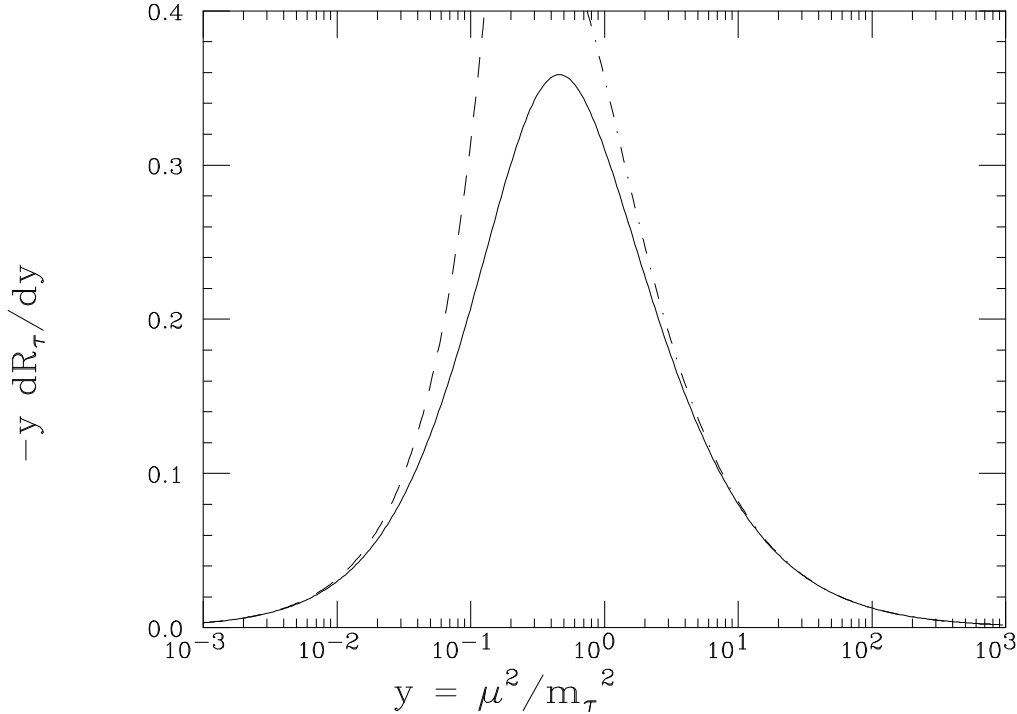}
\caption{Derivative of characteristic function for $\tau$ decay,
$\dot\cR_\tau$. Dashed and dot-dashed curves show the limiting
behaviour at small and large $y$, respectively.}
\label{fig_Rtau}
\end{figure}

The form of $\dot\cR_{\tau}(y)$ is shown by the solid curve
in Fig.~\ref{fig_Rtau}. At large $y$ the behaviour is
\beq
\dot\cR_{\tau}(y)= \frac 1 {5y}
\left(\ln y +\frac{107}{60}\right)
+{\cal O}\left(\frac{\ln y}{y^2}\right)\;,
\eeq
shown by the dot-dashed curve. At small $y$
\beq\label{cRtauas}
 \dot{\cR}_\tau(y) = 8\left(4-3\zeta(3)\right)y-18y^2
+\left(6\ln^2 y-12\ln y+\frac{32}{3}\right)y^3 
 \>+\> \ldots\,.
\eeq
(dashed).

The fact that
the non-analyticity of the expansion (\ref{cRtauas}) starts 
at order $y^3$ is again in accord with the absence of a
dimension-four contribution in the operator product expansion 
\citd{BBB}{SVZ}.
In this case the $\ln^2 y$ factor indicates a logarithmic
enhancement of the power-behaved correction to $R_\tau$, which
takes the form
\beq\label{RtauNP}
R_\tau^{\mbox{\scriptsize NP}}
\simeq -6\,(2\Apr_6\ln m_\tau^2 + 2\Apr_6-\App_6)/m_\tau^6\;.
\eeq

\subsection{\boldmath DIS structure function $F_2$}
We next consider the deep inelastic scattering process
in which a quark, with momentum $p$, is probed by a hard spacelike
photon of momentum $q$ ($q^2=-Q^2$, $ x=Q^2/2pq$).
Recall from Sect.~3 that in the case of
collinear singular quantities, such as DIS structure functions
(or their moments) and \ee fragmentation functions,
the characteristic function $\dot\cF(Q^2,\{x\};\mu^2)$
has a finite  $Q^2$-independent collinear limit,
$\dot\cF(Q^2,\{x\};0)\equiv P(x)$.
For small $\eps$ (inside kinematical limits),
\beq\label{dotFsing}
 \dot\cF(\eps) = P(x)  + \dot\cF_{reg}(\eps)\,,
\eeq
where $\dot\cF_{reg}(\eps)$ vanishes as a power of $\eps$ at $\eps\to 
0$,
as does $\dot\cF$ for the collinear safe case, see (\ref{Fsmalleps}).  
As a consequence, the function $\ddot\cF$ 
vanishes as a power for both $\eps\to 0$ and $\eps\to\infty$.
This quantity determines the scaling violation according to
Eq.~\re{scaviol}.  More precisely, defining the moments of the
non-singlet part of the structure function $F_2$ as
\beq\label{F2mom}
F_2(Q^2,N) = \int_0^1 x^{N-1} F_2(Q^2,x)\,dx\;, 
\eeq
the scaling violation is described by
\beq\label{DISscaviol}
\Gamma_N(Q^2)\equiv  Q^2\frac{\partial}{\partial Q^2} \ln F_2(Q^2,N)
\>=\> \frac{C_F}{2\pi}\int_0^\infty \frac{d\mu^2}{\mu^2}\> \at(\mu^2)
\> \ddot\cF(\eps=\mu^2/Q^2,N)\>.
\eeq

\paragraph{Evaluation of characteristic function.}
The characteristic function $\cF(\eps)$ is obtained by considering the 
amplitude for an incoming and outgoing quark and for an off-shell 
gluon of momentum $k$, either virtual or real.
For the combination of polarizations leading to
the $F_2$  structure function, 
the amplitude-squared for off-shell gluon emission assumes the 
following simple form (apart from the coupling $\as$ and the quark 
colour factor $C_F/2\pi$)
\beq\label{MDIS}
\cM_{2}=
\frac{y(1-x)-x\eps}{(y-x\eps)^2}
+\frac{2x(1-y)(1-\eps)}{(y-x\eps)(1-x)}
+\frac{y(1-x)-x\eps}{(1-x)^2}
+6x\frac{y^2(1-y)}{(y-x\eps)^2}\,,
\eeq
with
$ y = (pk)/(pq)$.
The last term in \re{MDIS} is the one contributing to the longitudinal 
structure function $F_L$ . When
$\eps\!=\!0$ the first and second terms are collinear singular for 
$y\! \to\! 0$, with coefficient equal to the quark splitting function
$P(x) = (1+x^2)/(1-x)$.

The real part of the characteristic function contributing to $F_2$
is obtained by integrating $y$ over the phase space region 
$\eps x/(1-x)< y <1$. 
Including the virtual contribution, which is now $\cV_s(\eps)$ given
by Eq.~\re{virts}, one finds that
\beq\label{FxDIS}
\cF(x;\eps) = \cFr(x;\eps)\>\Theta(1-x-\eps x) + 
\cV_s(\eps)\>\delta(1-x)
\eeq
with
\beq\label{FrDIS}\eqalign{
\cFr(x;\eps) &= \left[\frac {2(1-\eps)^2}{1-x}-(1+x)
+2(2+x+6x^2)\eps -2(1+x+9x^3)\eps^2\right]
\ln\left[\frac{(1-\eps x)(1-x)}{\eps x^2}\right] 
\cr&
-\frac{3+14\eps-15\eps^2}{2(1-x)} +\frac{\eps}{(1-x)^2}
+\frac{\eps^2}{2(1-x)^3}+\frac{x}{1-\eps x}
\cr& 
+1+3x+6(1-x)(1+3x)\eps-(8+9x+18x^2)\eps^2 \;.
}\eeq
The coefficient of $-\ln \eps $ is the quark splitting function 
$P(x)$, which is singular for $x\to 1$.
This singularity is regularized by including the 
virtual contribution. The $F_2$ quark structure function
to this order is thus (we suppress the standard
parton model factor $\sum_f e_f^2$) 
\beq\label{F2DIS}
F_2(x,Q^2)\>=\>\delta(1-x) \>+\> \frac{C_F}{2\pi}\int_0^\infty 
\;\frac{d\mu^2}{\mu^2}\, \at(\mu^2)\;\dot\cF(x;\eps)\;.
\eeq
The integral is convergent for $\mu^2 \to \infty$. In this limit 
the characteristic function $\cF(\eps)$ is given by the virtual 
contribution alone and we have
\beq
 \cF= -\frac2{3\eps}\left(\ln\eps+\frac{11}{6}\right) 
\>+\> \cO{\frac{\ln\eps}{\eps^2}}
\,.
\eeq

\begin{figure}
\vspace{9.0cm}
\includegraphics{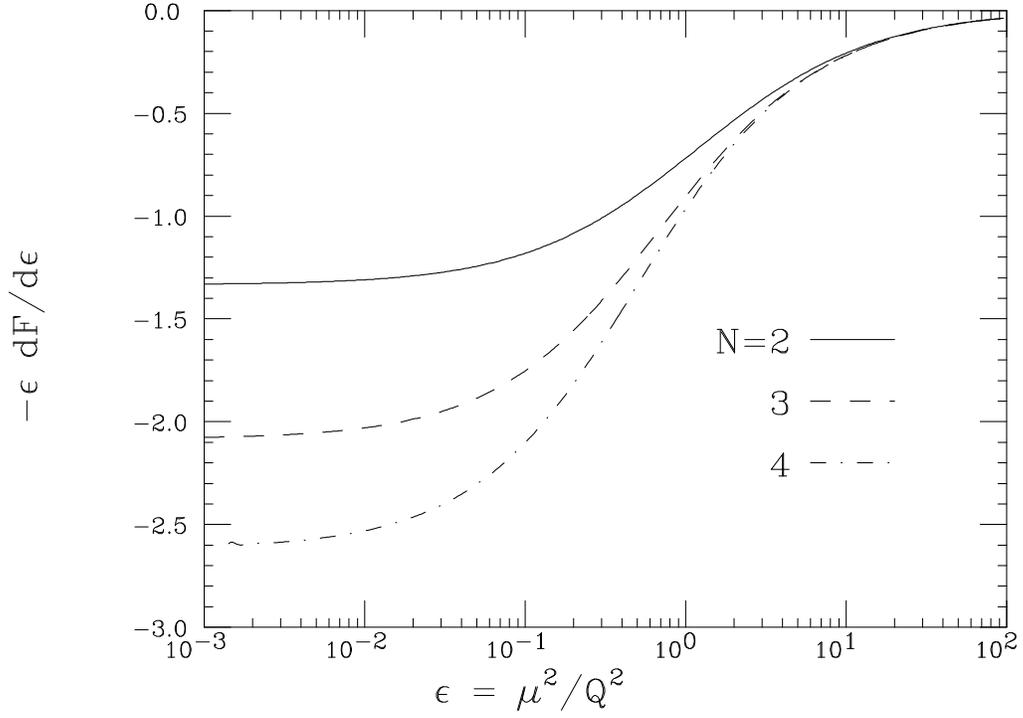}
\caption{Derivative of characteristic function for DIS, $\dot\cF_N$.}
\label{F2Nfig1}
\end{figure}

\begin{figure}
\vspace{9.0cm}
\includegraphics{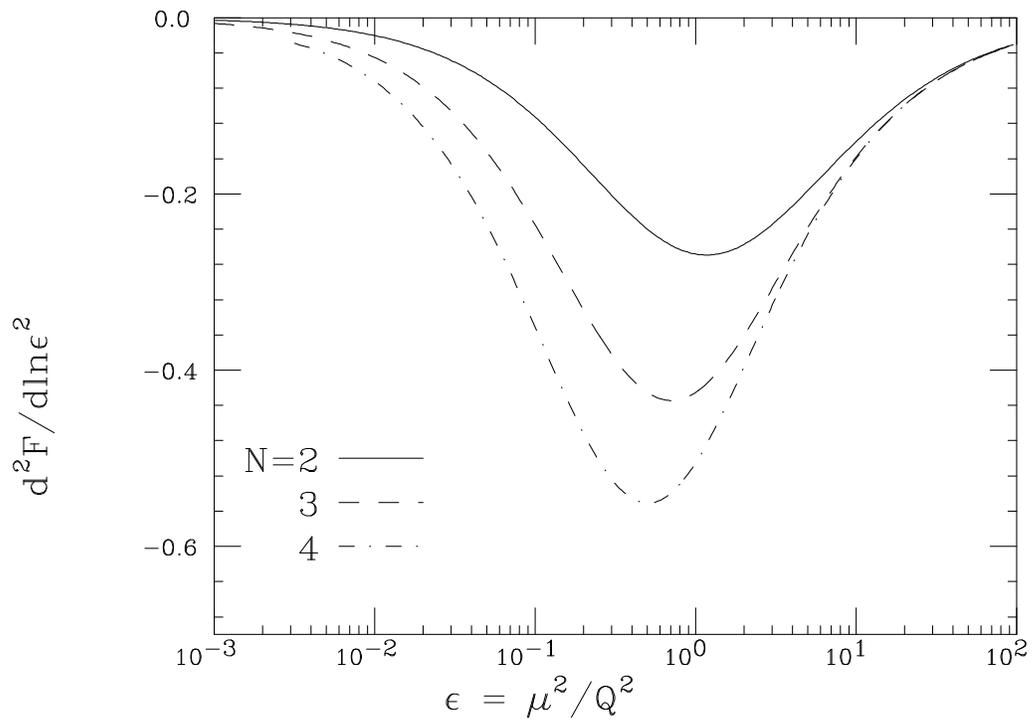}
\caption{Second derivative of characteristic function for DIS,
$\ddot\cF_N$.}
\label{F2Nfig2}
\end{figure}

In Figs.~\ref{F2Nfig1} and \ref{F2Nfig2} we show the first few moments
of $\dot\cF$ and $\ddot\cF$, for $N=2,3,4$. Note that the $N=1$
moment of $\dot\cF$ vanishes owing to the following
identity, which holds for any $\eps$,
\beq\label{Mvint}
\cV_s(\eps) = -\int_0^1 \cFr(x;\eps)\Theta(1-x-\eps x)\,dx\;.
\eeq
This means that the Adler sum rule is satisfied identically, that is,
it receives neither perturbative nor power corrections within our 
approach
(see \cite{AEM}).

\paragraph{Logarithmic scaling violation.}
The moments of the limiting function $P(x)$ in (\ref{dotFsing})
have the meaning of the anomalous dimensions
$\gamma_N$.\footnote{It should be noted that in higher orders
$P(x)$ will start to depend on the integration scale through
the effective coupling, $P \>=\> P(x;\at(\mu^2))$, while
remaining $Q^2$-independent.}
The usual perturbative scaling violation result for the
moments of the DIS structure function, namely, 
\beq
\Gamma_N(Q^2) \>=\> \gamma_N(\as(Q^2)) \>+\>
  \frac{d\, \as(Q^2)}{d\ln Q^2}\, 
\partder{\as}\,C_N(\as) \>,
\eeq  
is then reproduced as follows. 
Since the integrand in (\ref{scaviol}) is peaked at
$\mu^2=\bar{Q}^2\sim Q^2$, 
one obtains the leading contribution by substituting 
$\at(\mu^2)\to \as(\bar{Q}^2)$:
\beeq\label{sv0}
\Gamma_N^{(1)}&=& \as(\bar{Q}^2)\frac{C_F}{2\pi}
 \int_0^\infty \frac{d\mu^2}{\mu^2}
\> \ddot\cF(Q^2,N;\mu^2)\nonumber \\
&=& \as(\bar{Q}^2)\frac{C_F}{2\pi} 
\left[\, \dot\cF(Q^2,N;0)-\dot\cF(Q^2,N;\infty)\,\right]
= \as(\bar{Q}^2)\frac{C_F}{2\pi}P_N 
\eeeq
where $P_N$ is the corresponding moment of $P(x)$. 
Keeping the next term in the expansion for the running coupling,
$$
 \as(\mu^2)- \as(\bar{Q}^2) \>=\> 
\frac{d\, \as(\bar{Q}^2)}{d\ln \bar{Q}^2}\, 
 \ln\frac{{\mu}^2}{\bar{Q}^2} \>+\>\ldots,
$$
the first correction to (\ref{sv0}) can be derived as
\beq\label{sv1}
\Gamma_N -\Gamma_N^{(1)} \approx \frac{d\, \as(\bar{Q}^2)}{d\ln 
\bar{Q}^2} 
\frac{C_F}{2\pi}\int_0^\infty \frac{d\mu^2}{\mu^2} 
\>\ln\frac{\bar{Q}^2}{\mu^2}\> \ddot\cF(Q^2,N;\mu^2) = 
 \frac{d\, \as(\bar{Q}^2)}{d\ln \bar{Q}^2}\frac{C_F}{2\pi}c_N 
\;,
\eeq
where
\beq
c_N\>\equiv\>   \lim_{\mu^2\to0} 
\left\{\cF(Q^2,N;\mu^2)-\ln\frac{\bar{Q}^2}{\mu^2}\cdot P_N \right\}
=\cF_{reg}(Q^2,N;0)
\eeq
is the coefficient function, the value of which clearly depends on the
choice of the expansion scale $\bar{Q}^2$.
An obvious (first-order) identification is as follows,
\beq
  \gamma_N(\as) \Longleftrightarrow \as C_F P_N/2\pi\>, \qquad 
  C_N(\as) \Longleftrightarrow \as C_F c_N/2\pi \>.
\eeq
Notice that in the original expression (\ref{DISscaviol}) there is no
arbitrariness in the choice of the hard scale (scheme dependence),
which only emerges when one tries to evaluate the
area under the $\ddot\cF$ curves in Fig.~\ref{F2Nfig2} 
(weighted with $\at(\mu^2)$) in terms of a series expansion around
a given point $\bar{Q}^2$.

Another point that is clearly seen from Fig.~\ref{F2Nfig2}
concerns the actual hardness scale of the process. The
region from which the scaling violation rate receives its
main contribution shifts to lower momentum scales with
increasing $N$; in fact the peak is around
$\bar Q^2\sim Q^2/N$, or, equivalently, $Q^2(1-x)\approx W^2$.
To avoid confusion, let us stress that identification of $W^2$ with 
the
proper physical hardness scale is true for non-singlet structure 
functions
only. Therefore it should not be applied to the small-$x$ region,
which is dominated by the singlet contribution, the scale of which
does exceed $Q^2$ but by far less than the kinematically allowed limit 
$W^2\approx Q^2/x\gg Q^2$.
  
To obtain the power corrections we have to consider the small-$\eps$ 
limit of the characteristic function.

\paragraph{\boldmath Small-$\eps$ behaviour.}
When taking this limit, one has to exercise some care
in treating the singular functions and distributions in 
Eq.~(\ref{FrDIS}).
It follows from the identity \re{Mvint} that
for any test function $f(x)$ we have
\beq\label{Ffint}\eqalign{
\int_0^1\cF(x;\eps)\,f(x)\,dx &=
\int_0^1 \cFr(x;\eps)\Theta(1-x-\eps x)\left[f(x)-f(1)\right]\,dx \cr
&= \int_0^1 \cFr(x;\eps)\left[f(x)-f(1)\right]\,dx
+f'(1)\int_{1/(1+\eps)}^1 \cFr(x;\eps)(1-x)\,dx +\ldots\;.
}\eeq
Defining `+' and `++' prescriptions such that
\beq\eqalign{
  F(x)_+ &= F(x)-\delta(1-x)\int_0^1 F(z)\,dz\,,\cr
  F(x)_{++} &= F(x)_+ +\delta'(1-x)\int_0^1 F(z)(1-z)\,dz\,,
}\eeq
and recalling that
\beq
\int_0^1 \delta'(1-x)\,f(x)\,dx = f'(1)\;,
\eeq
we should therefore define the singular terms in
Eq.~(\ref{FrDIS}) at small $\eps$,
up to terms of order $\eps$, as follows:
\beq\label{smalleps}\eqalign{
  \frac{1}{1-x} &\to \frac{1}{(1-x)_+} +\eps\,\delta'(1-x)\cr
  \frac{\ln(1-x)}{1-x} &\to \left(\frac{\ln(1-x)}{1-x}\right)_+
+(\eps\ln\eps-\eps)\,\delta'(1-x)\cr
  \frac{\eps}{(1-x)^2} &\to \frac{\eps}{(1-x)^2_{++}}
+\eps\ln\eps\,\delta'(1-x)\cr
  \frac{\eps^2}{(1-x)^3} &\to -\eps\,\delta'(1-x)\;.
}\eeq
The small-$\eps$ behaviour of $\cF$ is thus of the form
\beq\label{Fxsmalleps}
\cF(x;\eps) =
-P(x) \ln\eps +c(x) +g(x)\,\eps\ln\eps +h(x)\,\eps
+\cO{\eps^2\ln \eps}\,,
\eeq
where
\bmini
 P(x) &=& \frac{2}{(1-x)_+} -(1+x) +\frac 32\, \delta(1-x)
= \left(\frac{1+x^2}{1-x}\right)_+ \\
 c(x) &=&  2\left(\frac{\ln[(1-x)/x^2]}{1-x}\right)_+
 -(1+x)\ln\left(\frac{1-x}{x^2}\right)+1+4x \nonumber \\
&& -\frac{3}{2(1-x)_+}-\frac 94\,\delta(1-x) \\
 g(x) &=& \frac{4}{(1-x)_+} -2(2+x+6x^2)
+9\,\delta(1-x)+\delta'(1-x) \\
 h(x) &=& -4\left(\frac{\ln[(1-x)/x^2]}{1-x}\right)_+
+2(2+x+6x^2)\ln\left(\frac{1-x}{x^2}\right) \nonumber \\
&& -\frac{9}{(1-x)_+}+\frac{1}{(1-x)^2_{++}}
+8+13x-16x^2-8\,\delta(1-x) -4\,\delta'(1-x)\;.
\emini
The corresponding formula in moment space is
\beq\label{FNsmalleps}
\cF_N(\eps) = -P_N \ln\eps
+c_N +g_N\,\eps\ln\eps +h_N\,\eps +\cO{\eps^2\ln \eps}\,,
\eeq
where
\bmini
 P_N &=& -2S_1 +\frac 32 -\frac 1N -\frac 1{N+1} \\
 c_N &=& \left(\frac 32 +\frac 1N +\frac 1{N+1}\right) S_1 + S_1^2 
-3S_2
 -\frac 94+\frac{2}{N}+\frac{3}{N+1} -\frac{1}{N^2} -\frac{1}{(N+1)^2} 
\\ 
 g_N &=& N +8 -\frac 4N -\frac{2}{N+1}-\frac{12}{N+2}-4S_1 \\
 h_N &=& \left(N +8 -\frac 4N -\frac{2}{N+1}-\frac{12}{N+2}\right) S_1
- 2S_1^2 +6S_2 \nonumber \\
&& -5N-3+\frac{3}{N+1}+\frac{2}{N+2} +\frac{4}{N^2}
+\frac{2}{(N+1)^2}+\frac{12}{(N+2)^2}
\emini
with
\beq\label{Sdefs}
  S_p = \sum_{j=1}^{N-1} \frac 1 {j^p}\;;\;\;\;\;
  S_1 = \psi(N)+\gamma_E = \ln N+\cO{1/N}\,,\;\;\;\;
  S_2 = \frac{\pi^2}{6}-\psi'(N)\,.
\eeq

\begin{figure}
\vspace{9.0cm}
\includegraphics{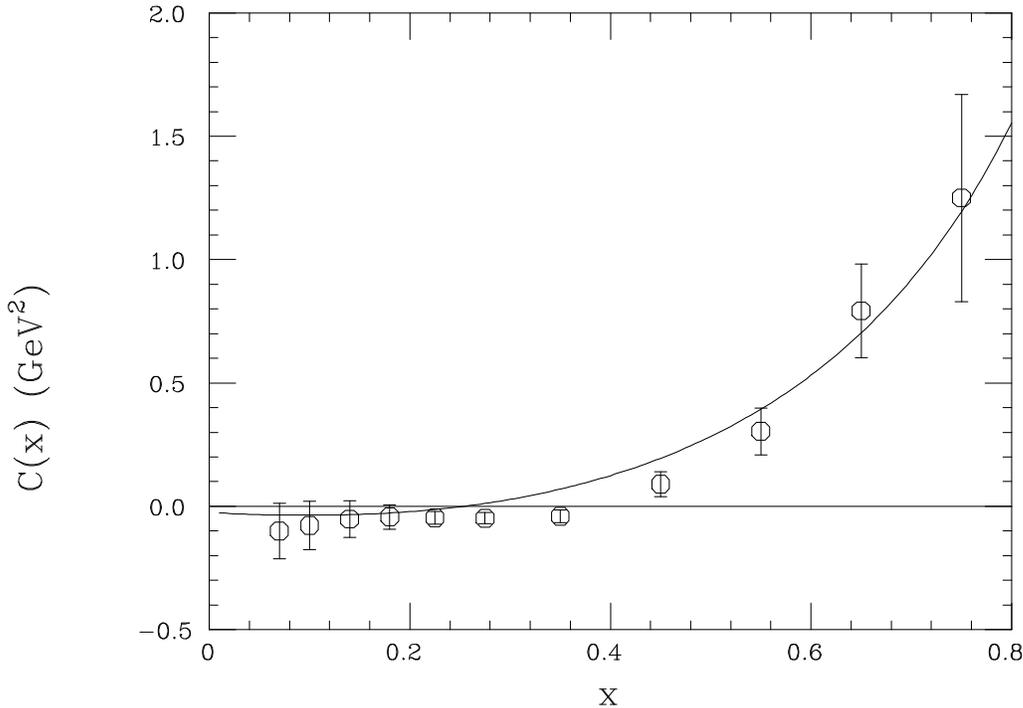}
\caption{Coefficient of higher-twist contribution to DIS.}
\label{fig_twist}
\end{figure}

\paragraph{Power corrections.}
From \re{Fxsmalleps} and \re{FNsmalleps}
we can now evaluate the dispersively-generated
power corrections to the usual (logarithmic)
scaling violation using Eq.~\re{DISscaviol}.
The leading non-analytic term of $\ddot\cF(\eps,N)$ for small $\eps$ 
is
$g_N\eps\ln\eps$, leading to a power-behaved contribution of the form
\beq\label{scviolNP}
 \frac{d}{d\ln Q^2}\; \ln F_2^{\mbox{\scriptsize NP}}(N,Q^2) 
\>=\> g_N \frac{\Apr_2}{Q^2}
\,.
\eeq
We shall not undertake detailed phenomenological studies in this 
paper,
but only indicate how the above result could be applied.  In the
analysis of DIS data, the non-perturbative contribution is normally
parametrized as a simple multiplicative correction of the form
\beq\label{F2twist}
F_2(x,Q^2) = \FPT_2(x,Q^2)\left(1+\frac{C(x)}{Q^2}\right)\;. 
\eeq
The `higher-twist'
coefficient $C(x)$ is then found to be a steeply
increasing function of $x$, as illustrated by the data points
in Fig.~\ref{fig_twist} \cite{VM}.

In our approach, the leading non-perturbative contribution is
of the form
\beq
F_2^{NP}(x,Q^2) = -\Apr_2\,g(x)\otimes \FPT_2(x,Q^2)/Q^2\;,
\eeq
where `$\otimes$' represents the convolution corresponding to a 
product
in moment space. Thus we predict a coefficient in Eq.~\re{F2twist}
with a weak (logarithmic) $Q^2$ dependence,
\beq\label{Ctwist}
C(x,Q^2) = -\Apr_2\,g(x)\otimes \FPT_2(x,Q^2)/\FPT_2\;.
\eeq

The curve in Fig.~\ref{fig_twist} shows the prediction of 
Eq.~\re{Ctwist}
at $Q^2=10$ GeV$^2$, assuming $A_2^\prime=-1$ GeV$^2$.
For $\FPT_2$ we have used the valence part of the MRSA
parametrization \cite{MRSA}.  We see that the form of the
observed higher-twist correction is well reproduced.  The steep
rise at large $x$ results mainly from the most singular term
$\delta'(1-x)$ in $g(x)$, which generates the behaviour
\beq\label{Ctwist2}
C(x,Q^2) \sim \Apr_2\,\frac{\partial}{\partial x}\ln \FPT_2(x,Q^2)
\eeq
at $x\to 1$. However, the less singular terms also play a significant
role in the region of $x$ shown in Fig.~\ref{fig_twist}.

\subsection{DIS sum rules}
\paragraph{Adler sum rule.}
As we have already mentioned above, due to the identity \re{Mvint}
the Adler sum rule (for a quark as a target)
\beq
   \int_0^1 dx \> F_2(Q^2,x) = 1
\eeq
does not acquire either perturbative or non-perturbative correction to 
this
order. 

\paragraph{Gross-Llewellyn-Smith sum rule.}
This sum rule concerns the first moment of the $F_3$ structure 
function.
For a target hadron $h$ one has
\beq
G(Q^2)=\int_0^1 dx [F^{\nu h}_3(Q^2,x)+F^{\bar\nu h}_3(Q^2,x)]
\;=\;G_0\;\int_0^1 dx F_3(Q^2,x)
\,,
\eeq
where $G_0=6B-2S$, with $B$ and $S$ the baryon number and strangeness 
of the hadron $h$, is the parton-model value. 
Here $F_3(Q^2,x)$ stands for the non-singlet structure 
function for lepton scattering by a single quark. 

According to our procedure we first compute the 
matrix element squared for massive gluon emission
corresponding to $F_3$. 
A simple calculation gives (apart from the coupling $\as$ 
and the colour factor $C_F/2\pi$)
\beq
\cM_d=\cM_3-\cM_2= 
{\frac {6\,yx-4\,x+2\,x\eps-2\,y}{
\left (y-x\eps\right )^{2}}}y^2  \,,
\eeq
where $\cM_2$ is given by (\ref{MDIS}).
Since the virtual contributions to $F_2$ and $F_3$ are identical, 
the difference between their characteristic functions is given by
\beq
\eqalign{
&
\cF_3(x,\eps)-\cF_2(x,\eps)=\int_{x\eps/(1-x)}^{1}dy\,\cM_d(x,y,\eps)
= -2\,\left(9\,x\eps-\eps-4\right ){x}^{2}\eps\ln \left({\frac 
{{x}^{2}\eps}
{(1-x\eps)(1-x )}}\right)
\cr&
-(1+x)-{\frac {2\,{x}^{2}\eps}{1-x\eps}} +  \eps\left 
(18\,{x}^{2}\eps+18\,{
x}^{2}+7\,x\eps-10\,x+5\,\eps-4-{\frac {2\,\eps}{\left (1-x\right 
)^{2}}}+{\frac {4-3\,\eps}{1-x}}\right )\;.
}
\eeq
This difference, determined by real emission only, is defined for 
$\eps \le (1-x)/x$ and vanishes at the phase space boundary as 
$$
\cF_3(x,\eps)-\cF_2(x,\eps)=
-(1+x)\left(\frac\eps{1-x}-\frac 1 x\right)^2+\ldots
\,.
$$
The first moment reads 
\beq\eqalign{
\cF_3(\eps,N\!=\!1) &= \int_{0}^{1/(1+\eps)}dx\, \cF_3(x) \cr
&=
-\frac 32
-{\frac {\epsilon}{2}}
+\epsilon\,
\left(\frac43 
+\frac56\,\epsilon\right)\ln{\frac{\epsilon+1}{\epsilon}}
-\left(\frac16 +\frac{2\eps}{3}\right) \left(
\epsilon^{-2} \ln(1+\epsilon)-\eps^{-1}+ \frac12 \right )
\,.
}
\eeq
It has the following small-$\eps$ behaviour, relevant for
power corrections: 
\beq
\cF_3(\eps,N\!=\!1)=
-\frac {3}{2} -\eps\left ( \frac43\,\ln\eps +\frac59 \right )
 - \eps^2\left (\frac 56\,\ln\eps-\frac {83}{72} \right)
+ \cO{\eps^3}\;.
\eeq
Thus the one-loop correction to the first moment of the quark 
structure 
function $F_3$ is
\beq\label{F3pred}
F_3(Q^2,N\!=\!1) = \frac{C_F}{2\pi}\int_0^\infty
\frac{d\mu^2}{\mu^2}\at(\mu^2)\dot \cF_3(\eps,N\!=\!1)
\,,
\eeq
where
\beq\label{dotF3}
\dot\cF_3(\eps,N\!=\!1)=
\frac12 
+ \frac {5\,\epsilon}{3} 
-\epsilon\,\left(\frac53\,\epsilon+\frac43\right )
\ln \frac {\epsilon+1}{\epsilon} 
- \frac {\left (2\,\epsilon+1\right )\ln (\epsilon+1)-\epsilon}
{3\,\epsilon^{2}}
\,.
\eeq
As usual (for a collinear-safe quantity) this function vanishes for 
both
large and small $\eps$. 
The area under the curve gives the well-known one-loop perturbative 
contribution
\beq
G^{\mbox{\scriptsize 
PT}}(Q^2)=G_0\;\left[\,1-\frac{3C_F}{4\pi}\as(Q^2) 
+\cO{\as^2(Q^2)}\, \right].
\eeq
We may now evaluate the dispersively-generated
power corrections, which are determined 
by the small-$\eps$ behaviour
\beq
 \dot\cF_3(N\!=\!1,\eps)= \eps\left(\frac43\ln\eps+\frac{17}9\right) 
+ \eps^2\left(\frac53\ln\eps-\frac{53}{36}\right) 
+ \mbox{regular terms}\,.
\eeq 
The two non-analytic terms generate power-behaved contributions
of the form
\beq\label{GLSNP}
G^{\mbox{\scriptsize NP}}(Q^2)\simeq G_0\left(
\frac 43\frac{\Apr_2}{Q^2}+\frac 53\frac{\Apr_4}{Q^4}\right)
\;.
\eeq

The estimated coefficient of $1/Q^2$ \cite{BK} used in a recent
analysis at $Q^2\sim 3\;\GeV^2$ by the CCFR collaboration 
\cite{harris}
is $-0.09\pm 0.05\;\GeV^2$. This implies a small negative value
of $\Apr_2$. However, at such a low value of $Q^2$ the $1/Q^4$
contribution could also be significant.

\subsection{\boldmath \ee fragmentation function}
Our procedure for \ee annihilation is similar to that presented for
deep inelastic scattering. 
The \ee fragmentation function $\tilde F(x,Q^2)$ is given in terms 
of the characteristic function 
\beq
\cF(x;\eps) = \cFr(x;\eps)\Theta(1-\eps-x) + \cV_t(\eps)\delta(1-x)
\eeq
where $\cFr$ is the real contribution from annihilation into
quark-antiquark-gluon,
\beq\eqalign{
\cFr(x,\eps) &= \left[2\frac{(1+\eps)^2}{1-x} -(1+x)-2\eps\right]
\ln\left[\frac{(x+\eps)(1-x)}{\eps}\right] \cr
&+\frac 12 (1+x)+\eps -\frac{3+4\eps+3\eps^2}{2(1-x)}
+\frac{\eps(1+\eps)}{(1-x)^2}+\frac{\eps^2}{2(1-x)^3}
+\frac{\eps}{x+\eps}\;,
}\eeq
and $\cV_t(\eps)$ is the timelike virtual correction given in
Eq.~(\ref{virtt}).
In this case we do not have an exact sum rule
of the form (\ref{Mvint}): instead
\beq\label{Fvint}
\cV_t(\eps) = \frac 32 -\int_0^1 \cFr(x;\eps)\Theta(1-\eps-x)\,dx
+\cO{\eps^2}\;.
\eeq
However, this is sufficient for our purposes, because it means that 
for
any test function $f(x)$ we can still write
\beq\eqalign{
\int_0^1\cF(x;\eps)\,f(x)\,dx &= \frac 32 f(1)\cr
& + \int_0^1 \cFr(x;\eps)\left[f(x)-f(1)\right]\,dx
+f'(1)\int_{1-\eps}^1 \cFr(x;\eps)(1-x)\,dx +\ldots
}
\eeq
up to terms of order $\eps$.  Using the dictionary of singular
terms (\ref{smalleps}), we thus obtain in this case the small-$\eps$
behaviour (\ref{Fxsmalleps}) with the same splitting function $P(x)$
but with $c(x)$, $g(x)$ and $h(x)$ replaced by
\beeq\label{Cghee}
\tilde c(x) &=&  2\left(\frac{\ln[x(1-x)]}{1-x}\right)_+
 -(1+x)\ln[x(1-x)]+\frac 12(1+x)\nonumber \\
&& -\frac{3}{2(1-x)_+}-\frac 94\,\delta(1-x) \nonumber \\
\tilde g(x) &=& -\frac{4}{(1-x)_+} +2 +\delta'(1-x) \nonumber \\
\tilde h(x) &=& 4\left(\frac{\ln[x(1-x)]}{1-x}\right)_+
-2\ln[x(1-x)] +\frac{1}{(1-x)^2_{++}}
+\frac 2x \nonumber \\
&& -5\,\delta(1-x) -4\,\delta'(1-x)\;.
\eeeq
The term $2/x$ in $\tilde h(x)$ needs some interpretation.
It results from setting
\beq
\ln\left(1+\frac{\eps}{x}\right) + \frac{\eps}{x+\eps} \simeq 
\frac{2\eps}{x}
\eeq
which is not valid unless $x\gg\eps$. Hence the expression for $\tilde 
h(x)$
is not to be used in the small-$x$ region, or for moments that are 
sensitive
to it. In particular, in moment space we have a result of the form
(\ref{FNsmalleps}) with $\cF_N=\frac 32+\cO{\eps^2}$ for $N=1$,
while for $N>1$
\bmini
\tilde c_N &=& \left(\frac 32 +\frac 1N +\frac 1{N+1}\right) S_1 + 
S_1^2 +3S_2
 -\frac 94+\frac{3}{2N}-\frac{1}{2(N+1)} +\frac{2}{N^2} 
+\frac{2}{(N+1)^2} \\ 
\tilde g_N &=& N -1 +\frac 2N +4S_1 \\
\tilde h_N &=& \left(N-1+\frac2N\right) S_1 + 2S_1^2 +6S_2 
-5N+\frac{2}{N-1} 
+ \frac4{N^2}\;.
\emini

For phenomenological applications, taking the same approach as for 
DIS,
we may parametrize the leading power correction to the fragmentation
function as
\beq
\tilde F(x,Q^2) = \tFPT(x,Q^2)\left(1+\frac{\tilde 
C(x,Q^2)}{Q^2}\right) 
\eeq
where
\beq\label{Cee}
\tilde C(x,Q^2) =  -\Apr_2\,\tilde g(x) \otimes
\tFPT(x,Q^2)/\tFPT(x,Q^2)\;.
\eeq

\begin{figure}
\vspace{9.0cm}
\includegraphics{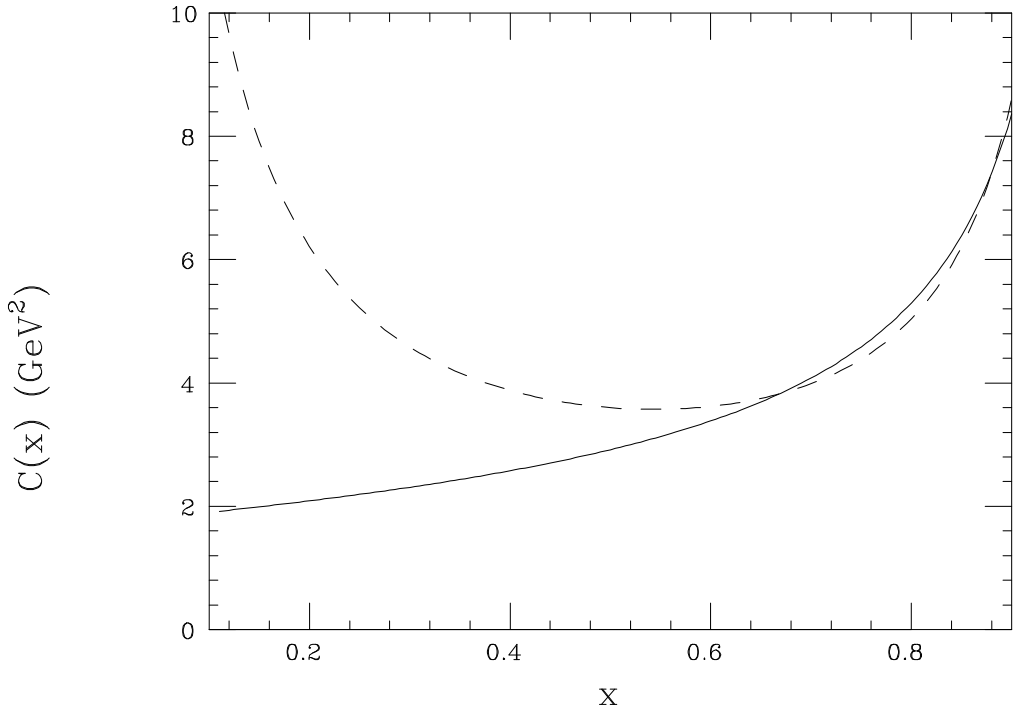}
\caption{Coefficient of power correction to \ee fragmentation 
function.}
\label{fig_had}
\end{figure}

Fig.~\ref{fig_had} shows the resulting prediction using the same
parameter value as for DIS, $A_2^\prime=-1$ GeV$^2$. 
For the perturbative contribution $\tFPT$ we used the
parametrization of the light quark fragmentation function at
$Q=22$ GeV given in Ref.~\cite{ALEPHfrag}. The coefficient function
$\tilde C$ is larger than that for DIS, but qualitatively similar
in form. The values of $Q^2$ in \ee annihilation being much larger,
the predicted power correction is probably too small to be detectable.

In Ref.~\cite{ALEPHfrag}, the non-perturbative contribution to the
fragmentation function was parametrized as a small shift in the
value of $x$,
\beq\label{xshift}
x\to x+h_0\left(\frac 1Q -\frac 1{Q_0} \right)
\eeq
where $h_0=-0.14\pm 0.10$ GeV. However, a parametrization with
$1/Q^2$ in the place of $1/Q$ was also found to be acceptable,
the magnitude of the correction being practically
consistent with zero. A small shift in $x$ is clearly equivalent
to a correction proportional to $\tilde F'$, 
the $x$-derivative of $\tilde F$,
as expected from the $\delta'$ term in $\tilde g(x)$.
As shown by the dashed curve in Fig.~\ref{fig_had}, a correction
of this form is similar to our prediction at large $x$.
The dashed curve corresponds to $\tilde C(x)=-0.35 \tilde F'/\tilde 
F$,
which, at 22 GeV, would be equivalent to $h_0\simeq -0.016$
in Eq.~\re{xshift}.

\subsection{Event shape variables}
The analysis of power corrections to event shapes in \ee final states
proceeds in the same way as for hard cross sections, except that in
this case the observables are constructed in such a way that there is 
no
virtual contribution.  Thus for the mean value of some generic shape
variable $y$, defined so as to vanish in the two-jet limit, we have
the characteristic function
\beq\label{cFshape}
\cF(\eps) = \cFr(\eps) = \int dx\,d\bar x\,\cM_{ee}(x,\bar x,\eps)
\,y(x,\bar x,\eps)\,\Theta[(1-x)(1-\bar x)-\eps]\,
\Theta[x+\bar x-1+\eps]\;,
\eeq
where $\cM_{ee}$ is the matrix element given in Eq.~(\ref{eeM}).

\paragraph{Thrust.}
In the case of the thrust variable $T$, for example, we define
$y = 1-T$, which vanishes in the two-jet limit, and
\beq\label{yTdef}
y(x,\bar x,\eps) = 
\min\{(1-x),(1-\bar x),(1-\sqrt{x_g^2-4\eps})\}\;.
\eeq 
where $x_g=2-x-\bar x$.
As before, the quantity to be inserted in the fundamental equation
(\ref{WDM}) is $\dot\cF$, the logarithmic derivative of the
function (\ref{cFshape}).

\begin{figure}
\vspace{9.0cm}
\includegraphics{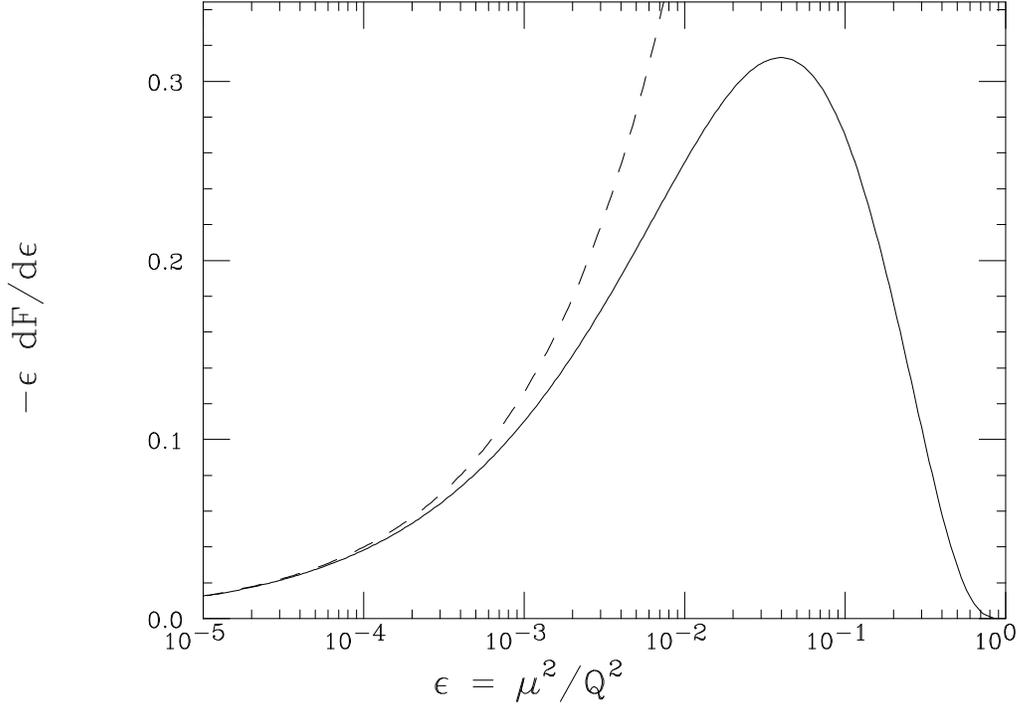}
\caption{Derivative of characteristic function for mean value of 
thrust.}
\label{fig_thrust}
\end{figure}

We see from Fig.~\ref{fig_thrust} that $\dot\cF$ for the mean thrust
decreases much more slowly at small $\eps$ than the corresponding 
function
for the quantities studied earlier.
In fact the behaviour at small $\eps$ is
\beq\label{cFTsmalleps}
\dot\cF(\eps)\sim 4\sqrt{\eps} + \cO{\eps\ln^2\eps}\;,
\eeq
which is shown by the dashed curve.  Thus the leading term
is non-analytic and of square-root type. It follows that the
leading power correction to $\lrang{1-T}$ will be of order
$1/Q$, as observed \cite{hadro}.

It is easy to see how the $\sqrt{\eps}$ behaviour arises. It comes
from the contribution to the derivative from the first theta-function
in Eq.~(\ref{cFshape}), \ie, from the soft phase-space boundary, which
gives
\beeq
\dot\cF(\eps)&\sim& \eps \int dx\,d\bar x\,\cM_{ee}(x,\bar x,0)
\,\min\{(1-x),(1-\bar x)\}\,\delta[(1-x)(1-\bar x)-\eps]\,
\Theta[x+\bar x-1]\nonumber \\
&\sim& 4 \int_{1-\sqrt\eps}^1 dx \>=\> 4\sqrt{\eps}\;.
\eeeq
Notice that we obtain in this approximation
\beq\label{VEV1T}
\lrang{1-T} \sim  \frac{C_F}{2\pi}
\int_0^1 dx\int_{1-x}^1 d\bar x\,\at[(1-x)(1-\bar x)Q^2]
\,\cM_{ee}(x,\bar x,0)\,\min\{(1-x),(1-\bar x)\}\;,
\eeq
which corresponds to the result obtained in Ref.~\cite{DW}:
the argument of $\at$ is the maximum gluon virtuality,
which is equal to its transverse momentum. 

In terms of the non-perturbative parameters \re{adefs},
the power correction to $\lrang{1-T}$ is
given by the $p=\half$ moment of $\delta\at$ as
\beq\label{Ta1Q}
\lrang{1-T}^{\mbox{\scriptsize NP}}
\simeq 4\frac{A_1}{Q}\;.
\eeq
The analysis performed in Ref.~\cite{DW} showed that this
gives a good description of the data for
\beq\label{A1val}
A_1\simeq 0.25\,\GeV.
\eeq

In Ref.~\cite{BB}, the same $1/Q$ dependence as
in \re{Ta1Q} was obtained, but with a different
coefficient. The difference arises from the
normalization factor in the definition of the thrust:
we have normalized to the sum of the final-state energies,
whereas in \cite{BB} the sum of momenta is used, corresponding
to inserting a factor of
\beq\label{BBfac}
2/\left(2-x_g+\sqrt{x_g^2-4\eps}\right)
\eeq
into Eq.~\re{yTdef}. Even though the thrust as usually
defined is normalized to the sum of final-state momenta,
the factor \re{BBfac} should not be included,
because it corresponds to finding a real massive
gluon in the final state. The massive gluon in
the calculation always decays into massless quarks
or gluons, and so the sum of the final-state momenta
should be set equal to the sum of the energies.

As mentioned in Sect.~3, the dispersive method applies
directly to quantities that are fully inclusive with respect to
gluon branching. Then the entire effect of branching is to make the
coupling run.  However, event shapes are sensitive to the
structure of the final state, and branching may lead to a different
value of the observable. In the case of thrust, for example, the
value is unchanged only if the products of gluon branching
fall into the same hemisphere, which includes quasi-collinear
branching and therefore gives the dominant contribution to
$\lrang{1-T}\sim \at/\pi$. Branching into opposite hemispheres,
corresponding to a genuine four-parton contribution to the
thrust, gives a correction $\sim (\at/\pi)^2$. Both give rise
to $1/Q$ power terms \cite{NS}. The extent to which terms
of higher order in $\at$ affect the magnitude of the
power-behaved contribution remains to be established
phenomenologically. The result \re{A1val} suggests a
typical value of the effective coupling in the small
momentum region such that $\at/\pi\sim 0.2$. This
gives grounds for optimism that higher-order terms
may be controllable.

Similar results to \re{Ta1Q}, with different coefficients
of $A_1/Q$, may be obtained for the mean values of a variety
of \ee and DIS final-state event shapes \citd{DW}{CDW}.

Since the presence of a $1/Q$ correction introduces a large
non-perturbative contribution, one would prefer for some
purposes to define event shape variables for which the
predicted coefficient of $1/Q$ is zero. Such variables
should be more suitable for testing perturbative predictions
and for measuring $\as$.  As suggested in \cite{hadro}, one
can find linear combinations of variables such that the
predicted $1/Q$ terms cancel, at least in the mean value. 
In addition, there are shape variables for which such
terms vanish because the small-$\eps$ behaviour of
the characteristic function is not of the square-root type.

\paragraph{Three-jet resolution.}
An example of a shape variable without a leading-order
$1/Q$ correction is the mean value of the
three-jet resolution variable
$y_3$, defined according to the Durham or $k_\perp$
algorithm \cite{Duralg}.
In lowest order, in the region $x>\bar x$, we have
\beq
y_3 = \min\{y_{q \bar q},y_{\bar q g}\}
\eeq
where
\beeq\label{yijD}
y_{q \bar q} &=&
\half\bar x^2\,(1-\cos\theta_{q \bar q})\nonumber \\
y_{\bar q g} &=&
\half\min\{\bar x^2,x_g^2\}(1-\cos\theta_{\bar q g})\;,
\eeeq
with
\beeq
\cos\theta_{q \bar q} &=& \frac{1-x_g-\eps}{x\,\bar x}
\nonumber \\
\cos\theta_{\bar q g} &=& \frac{x_g-2(1-x-\eps)/\bar x}
{\sqrt{x_g^2-4\eps}}\;.
\eeeq

\begin{figure}
\vspace{9.0cm}
\includegraphics{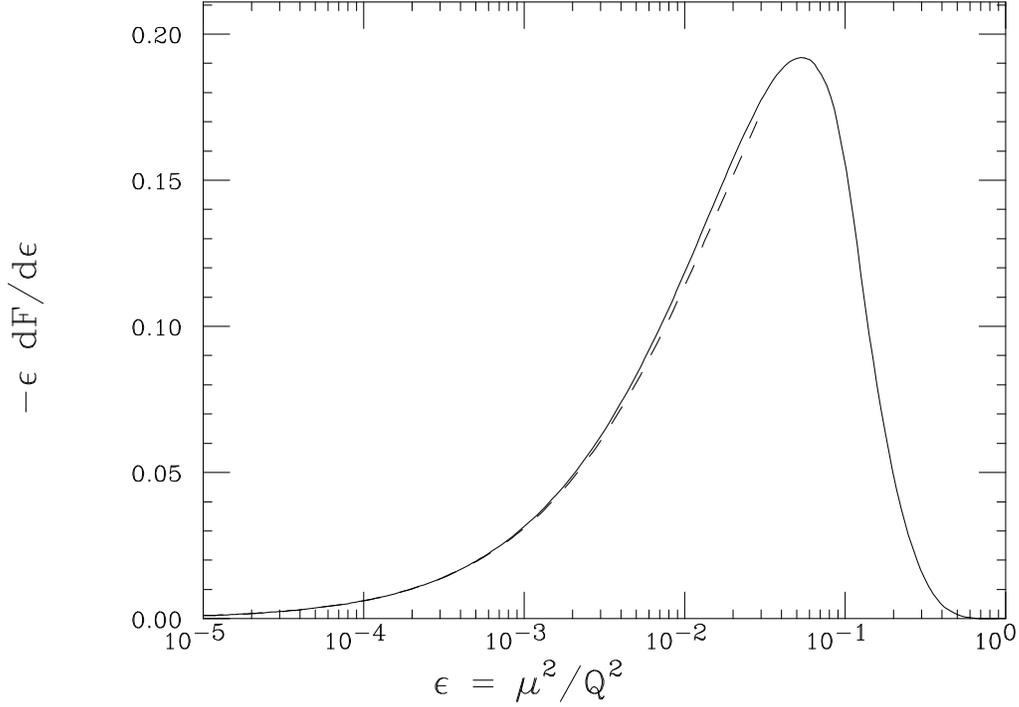}
\caption{Derivative of characteristic function for mean value of 
$y_3$.}
\label{fig_dury3}
\end{figure}

The resulting behaviour of $\dot\cF$ for the mean
value of $y_3$ is shown in Fig.~\ref{fig_dury3}.
The small-$\eps$ behaviour in this case is
\beq\label{cFysmalleps}
\dot\cF(\eps) \sim \eps\ln^2\eps + \cO{\eps\ln\eps}\;.
\eeq
It follows that the leading power correction to the mean value of
$y_3$ should be of order $\ln Q/Q^2$.
We find numerically that the non-leading logarithms are
such that
\beq\label{cFynum}
\dot\cF(\eps) \simeq \eps(\ln^2\eps + 3\ln\eps + 4)\;,
\eeq
as shown by the dashed curve.
 In terms of the moment integrals \re{adefs}, we therefore expect
\beq\label{y3NP}
\lrang{y_3}^{\mbox{\scriptsize NP}}
\simeq -(2\Apr_2 \ln Q^2 -3\Apr_2 -\App_2)/Q^2\;.
\eeq

If the three-jet resolution is defined instead according to the JADE
algorithm, then Eqs.~\re{yijD} become
\beeq\label{yijJ}
y_{q \bar q} &=&
\half x\,\bar x\,(1-\cos\theta_{q \bar q})\nonumber \\
y_{\bar q g} &=&
\half \bar x\,x_g\,(1-\cos\theta_{\bar q g})\;.
\eeeq
In this case the behaviour at small $\eps$ is of
square-root type, leading to a $1/Q$ power correction.
This is probably why the Durham algorithm has been found to
require smaller non-perturbative corrections \cite{Siggi}. 

A final point to be noted from Figs.~\ref{fig_thrust} and
\ref{fig_dury3} is that the characteristic scale of hardness
for event shapes, as typified by the peak of the characteristic 
function,
lies far below $Q^2$, at $\bar Q^2\sim 0.05 Q^2$. This is a general
feature of quantities which probe final-state structure in time-like
processes, such as event shapes and \ee fragmentation functions, and
could explain why fixed-order fits to such quantities favour small
scales and large values of the coupling.

\subsection{\boldmath Drell-Yan process and $K$ factor}

\begin{figure}
\vspace{9.0cm}
\includegraphics{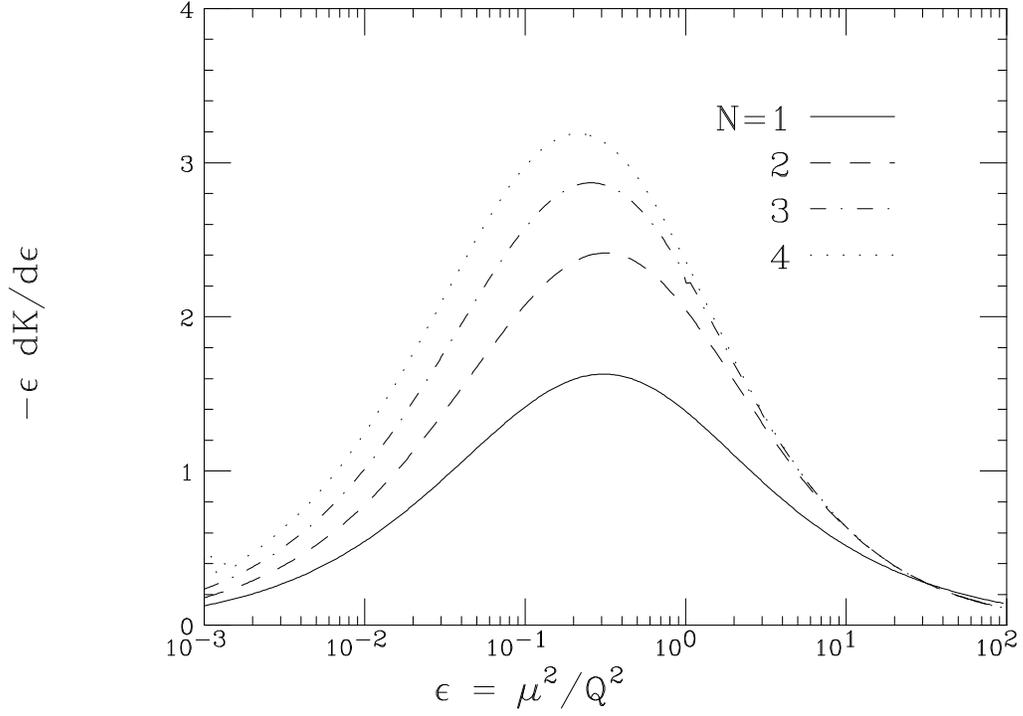}
\caption{Derivative of characteristic function for moments of the
Drell-Yan K-factor, $\dot\cK_N$.}
\label{fig_DYmom}
\end{figure}

The matrix element squared for the Drell-Yan (DY) process with
emission of an off-shell gluon  ($\mu^2>0$) is given by 
\beq
M_{DY}=
2{\frac 
{a+{\tau}^{2}a+2\,\tau\eps-2\,{\tau}^{2}{a}^{2}+2\,{\eps}^{2}\tau+
{\eps}^{2}{\tau}^{2}a}{\tau \,\left (\eps+a\right)^{2}}}
\,,
\eeq
where $\eps={\mu^2}/{Q^2}$, $\tau=Q^2/s$ and $a={q^2}/{Q^2}$ with $q^2$ 
the momentum integration variable. 
The phase space integration is given by
$$
d\Phi= da \frac{\tau}{\sqrt{R}}\;\Theta(R)
\,,
$$
with
$$
R=A-4\,{\tau}^{2}a\,,
\;\;\;\;\;
A=
1-2\,\tau+{\tau}^{2}
-2\,\tau\eps-2\,{\tau}^{2}\eps
+{\eps}^{2}{\tau}^{2}
\,.
$$
The real part of the DY characteristic function is then 
\beq
\cFr_{DY}(\eps,\tau)=\int_0^1 d\Phi\,M_{DY}\;
=
-4\,\sqrt A +4\, \tanh^{-1}
\left(\frac {\sqrt A}{1-\tau-\tau\eps} \right)
\frac{{\tau}^{2 }+1+2\,{\tau}^{2}\eps+{\tau}^{2}{\eps}^{2}}
{1-\tau-\tau\eps }
\,.
\eeq
We introduce the ``rapidity'' variables 
\beq\label{kin}
 \tau = \frac{c+\sqrt{c^2-1}}{2\left(\cosh\eta+c\right)} \>,
\qquad c=\frac{1+\eps}{2\sqrt{\eps}}
\,,
\eeq
with the phase space
$$
 d\tau = \frac{c+\sqrt{c^2-1}}{2}\,\frac{\sinh\eta\,d\eta}
{\left(\cosh\eta+c\right)^2}\>,
\qquad \tau_{\max}= \frac{c+\sqrt{c^2-1}}{2(1+c)} = 
\frac1{(1+\sqrt{\eps})^2}
\>.
$$
One then has the simple form
\beq
\cFr_{DY}=
-{\frac {4\,\sinh\eta}{\cosh\eta + c}}+4\,
\eta\,\left( 1+{\frac {c}{\cosh\eta }} \right)
\left( 1+ \frac {c^2} 
{\left( \cosh\eta+c \right)^2} \right )
\,.
\eeq
The complete characteristic function is obtained by including 
the virtual contribution. In moment space one has 
\beq
\cF_N^{DY}(\eps) =\int_0^{\tau_{\max}} d\tau\> \tau^{N-1} \>
\cFr_{DY}(\eps,\tau)\>+\>\cV_t(\eps)\,.
\eeq
This function is given to order $\eps$ in Appendix A, Eq.~\re{cFDYN}.
The DY distribution moments are then given by
\beq\label{DY}
\sigma_N^{DY}(Q^2)
=
\frac{C_F}{2\pi}\int_0^\infty \frac{d\mu^2}{\mu^2}
\>\at(\mu^2)\>
\dot \cF_N^{DY}(\eps) \,,
\eeq
and the moments of the Drell-Yan $K$ factor are
\beq\label{Kfact}
K_N(Q^2)
= \frac{C_F}{2\pi}\int_0^\infty \frac{d\mu^2}{\mu^2}
\>\at(\mu^2)\>
\dot \cK_N(\eps) \,,
\eeq
where the characteristic function is 
\beq\label{cKNdef}
\cK_N(\eps) = \cF_N^{DY}(\eps) - 2 \cF_N^{DIS}(\eps) 
\eeq
with $\cF_N^{DIS}(\eps)$ the corresponding moment of the DIS
characteristic function \re{FxDIS}.
As is well known, the combination \re{cKNdef} is collinear safe since 
the collinear singularities in the Drell-Yan and DIS terms cancel.
Its logarithmic derivative $\dot\cK_N(\eps)$ vanishes both for large
and small $\eps$, as shown for the first four moments in
Fig.~\ref{fig_DYmom}.  At small $\eps$, $\cK_N(\eps)$ has the form
\beq
\cK_N(\eps)=\hat c_N+\eps(\hat f_N \ln^2\eps +
 \hat g_N \ln\eps +\hat h_N) +\cO{\eps^2\ln^2\eps}\;,
\eeq
(see Appendix A, Eq.~\re{cKNeq}), 
which shows that, apart from logarithmic enhancement,
the power correction to the Drell-Yan cross section is of order 
$1/Q^2$,
as in DIS, in agreement with the result of \cite{BB}.

The expansion of $\dot\cK_N(\eps)$ at small $\eps$ has the form
\beq
\dot\cK_N(\eps)=\eps\left[\check f_N (\ln\eps+2S_1)^2+
\check g_N (\ln\eps+2S_1)+\check h_N \right]
\;(1+\cO{\eps})
\,,
\eeq
where at large $N$, corresponding to $\tau\to 1$, we have
\beq
\check f_N \sim N\;,\;\;\;\;
\check g_N \sim -2N^2\;,\;\;\;\;
\check h_N \sim 4N^2\;.
\eeq
Thus the expansion parameter at large $N$ becomes $\eps N^2$,
and for $\dot\cK_N$ one has approximately
\beq
\dot\cK_N(\eps)=2\eps N^2\left(\ln\frac{1}{\eps N^2}+2\right)
\left(1+\cO{\frac{\ln(\eps N^2)}{N}}\right)\;,
\eeq
where we have used
\beq
\ln\eps + 2S_1 \sim \ln(\eps N^2)\;.
\eeq
Correspondingly, at large $\tau$ the power correction to the $K$ 
factor
may be expressed in terms of the non-perturbative integrals \re{adefs} 
as
\beq
K^{\mbox{\scriptsize NP}}(\tau) = -\frac{2\Apr_2}{(1-\tau)^2Q^2}\;.
\eeq
This means that the coefficient of the power term increases much more 
rapidly
as $\tau\to 1$ than does the DIS coefficient $C(x)$ as $x\to 1$
in Eq.~\re{F2twist}.

\subsection{Summary}

\begin{table}\begin{center}
\begin{tabular}{|c|c|c|l|c|}
\hline
& & & & \\
 Process & Quantity & Power & Coefficient (first order) & Eq. \\ 
& & & & \\
\hline
& & & & \\
 \ee & $R$ & $Q^{-6}$ & $2 \Apr_6$ & \re{Rpc1} \\
& & & & \\
$\tau$ & $R_\tau$ & $m_\tau^{-6}$ & $ -12\Apr_6 L -12\Apr_6+6\App_6$&
\re{RtauNP} \\
& & & & \\
 DIS & $F_2$ & $Q^{-2}$ & $-g \Apr_2\otimes\FPT_2/\FPT_2$ & 
\re{Ctwist} \\
& & & & \\
 DIS & GLS & $Q^{-2}$ & $\frac43 \Apr_2 $ & \re{GLSNP} \\
& & & & \\
 \ee & $\tilde F$ & $Q^{-2}$ & $- \tilde g \Apr_2
\otimes \tFPT/\tFPT$ & \re{Cghee} \\
& & & & \\
\ee & $\lrang{1-T}$ & $Q^{-1}$ & $4 A_1$ & \re{Ta1Q} \\
& & & & \\
\ee & $\lrang{y_3}$ & $Q^{-2}$ & $-2 \Apr_2 L
+3\Apr_2 +\App_2$ & \re{y3NP} \\
& & & & \\
 DY & $K$ & $Q^{-2}$ & $
2\hat f \Apr_2 L -(2\hat f +\hat g) \Apr_2
-\hat f \App_2$ & \re{cKNeq} \\
& & & & \\ \hline
\end{tabular}
\caption{Summary of power-behaved contributions.}
\end{center}\end{table}

We summarize in Table 1 the results obtained for the leading
power-behaved corrections to the quantities discussed
above. In each case the predictions are expressed in terms
of the effective coupling moment parameters
given in Eq.~\re{adefs}, and $L=\ln Q^2$
($\ln m_\tau^2$ in the case of $R_\tau$).

At present no systematic phenomenological tests of these
predictions have been performed and the values of the moment
parameters are correspondingly uncertain. The parameter $A_1$
specifies the lowest-order $1/Q$ contributions
to a wide range of event shapes in \ee (and DIS) final
states, of which we show thrust only as an example.
A value of $A_1\sim 0.25\;\GeV$ is suggested by the
$1\;\GeV/Q$ power-behaved contribution seen in the data
on $\lrang{1-T}$.

As discussed in Sect.~4.2, the $1/Q^2$ contributions to
the structure function $F_2$ in DIS
appear consistent with a log-moment parameter value
$\Apr_2 \sim -1\;\GeV^2$.
The power-behaved contribution to the Gross-Llewellyn-Smith
sum rule for $F_3$ may suggest a smaller value.
In either case the corresponding contribution to
\ee fragmentation functions is probably too small to be observed.
The other $1/Q^2$ effects listed, for the mean three-jet resolution
in \ee final states and the $K$-factor in the Drell-Yan process,
show a logarithmic enhancement and therefore involve also
the additional log-moment parameter $\App_2$.

\mysection{Conclusions}

In this paper we have started to explore the possibility that
the concept of the QCD running coupling can be extended, 
in a process-independent way, down to small momentum scales, 
at least in an effective sense. At low scales confinement physics
dominates, undermining the very possibility of applying the language 
of
quark and gluons, the only language we are able to use at the present
level of our understanding of QCD. 
The hope is that, in spite of all the richness and complexity of the
spectrum of hadrons, the bulk of the non-perturbative physics 
will reveal itself in a smooth way in sufficiently inclusive
observables, and that it can be taken into account by extending the
notion of the perturbative QCD coupling. The coupling $\at$ thus
defined is intended to measure the effective interaction strength
even at large distances, allowing the extension 
of the quark-gluon language beyond its original domain of 
applicability. 

In an inclusive observable $F$, confinement reveals itself via 
departure 
of the measured value from the perturbative prediction $\FPT$
at decreasing values of the relevant hard scale of the process, $Q^2$. 
At sufficiently large $Q^2$, $\FPT$ varies slowly (logarithmically) 
with
$Q^2$, since it is given by a finite-order expansion in $\as(Q^2)$. 
The departure is typically power-behaved (see Table 1) 
\beq\label{pCk}
\FNP\>=\>
F(Q^2)-\FPT(Q^2) \simeq \left[\, C_1A_{2p}+C_2A_{2p}'+C_3A_{2p}''
                      \,\right]\cdot{Q^{-2p}}\,,
\eeq
where the dimensionless coefficients $C_k$ are at most linear in $\ln Q^2$.  
In this paper we have developed a method of calculating 
the process-dependent exponents $p$ and coefficients $C_k$ for a
wide variety of hard observables which do not involve gluons in the
Born approximation (quark-dominated processes).
The dimensionful parameters $A_{2p}$, $A_{2p}'$ and $A_{2p}''$
are universal and
are given by (log-)moment integrals of the effective coupling
modification $\delta\at$ in the small momentum region.
These are new phenomenological
parameters which should be studied experimentally. 
We have computed the corresponding power-behaved
contributions to first order in $\delta\at$ .

Our method is based on using a dispersion relation 
for the running coupling $\as$. 
The dispersive machinery then leads to the basic representation \re{WDM}
for the observable $F$ in terms of an integral over $\mu^2$ of the
product of a characteristic function $\dot\cF$ and the effective 
coupling $\at(\mu^2)$. The latter is related to the standard $\as$ by the 
operator equation
$$
 \at(\mu^2)= \frac{\sin(\pi\cP)}{\pi\cP}\, \as(\mu^2)\>, \quad 
 \cP=\mu^2\frac{ d}{d\mu^2}\>.
$$
This means that $\at$ and $\as$ are practically equivalent in the
perturbative region. At low scales, $\at$ and $\as$ may differ
substantially from each other and from expectations based on
the perturbative $\beta$-function. However, we have argued that
the non-perturbative modifications to $\at(\mu^2)$ at low $\mu^2$
should be such that no power corrections to $\as$ are generated
at higher scales. It follows that the $\mu^{2p}$-moments of these
deviations have to vanish for the first few integer values of $p$. 
Thus in order to generate power corrections, $\dot\cF$ must contain
terms which are non-analytic in $\mu^2$. 
    
The characteristic function $\dot\cF$ is specific to a given observable
and can be analysed by Feynman diagram techniques. We have computed 
the characteristic functions for a number of observables in various 
processes to first order in $\at$. 
In this order, $\dot\cF$ is a function of the ratio $\eps=\mu^2/Q^2$ 
and is given by the one-loop diagrams with a gluon of mass equal 
to the dispersive variable $\mu$, $0\le \mu \le \infty$. The leading 
non-analytic term in the $\eps\to0$ behaviour of $\dot\cF$ determines 
the power $p$ and coefficients $C_k$ in Eq.~\re{pCk}.

Let us stress again that we work with a massless gluonic quantum field.
However, in the characteristic function the dispersive variable $\mu$ 
enters as a gluon mass in Feynman denominators and the phase space.
The role of a small gluon mass as a trigger for long-distance
contributions to hard processes has been recognized and exploited 
in the recent literature, see \citq{hadro}{BB}{BBB}{BBZ}.

An attractive feature of the dispersive method is that for a given 
process it suffices to compute a single function to obtain power 
corrections to all associated observables. For example, the matrix 
element squared for \ee annihilation into massless \qq and a massive 
gluon gives the dispersively-generated power corrections to $R_\ee$,
to the non-singlet fragmentation function, and to all event shapes
(thrust, $y_3$, etc).  The results obtained for these and other hard
process observables are summarized in Table 1.

As emphasised in Table 1, our calculation of the coefficients of
power-behaved terms is first-order in the effective coupling $\at$.
The accuracy of the first-order estimate will be reasonable if
$\at/\pi$ happens to be numerically sufficiently small
in the important kinematic region.  This is a question that can
only be answered by comparison with experiment. The same qualification
applies even more strongly to event shapes which, by
construction, are sensitive to final state structure and not
fully inclusive. Multiparton contributions to such quantities are
higher-order in $\at$, and may or may not be suppressed, depending
on the typical value of $\at/\pi$.

On the topic of higher-order contributions,
we would like to end with the following remark.
A characteristic feature of the leading-order power-behaved terms
that we have calculated is that they are enhanced near the boundary of
phase space, where the additional gluon is soft.  One thus observes
the following general pattern of enhancement of power terms and the 
corresponding phase space boundaries:
$$\eqal2{
 \mbox{DIS coefficient functions}= 
& \frac{N}{Q^2}\L^2  &\Longrightarrow {(1-x)Q^2}\sim \L^2 \cr
 \mbox{DY $K$-factor}= 
& \frac{N^2}{Q^2}\L^2 &\Longrightarrow{(1-\tau)^2Q^2}\sim \L^2 \cr
 \mbox{differential thrust distribution}= 
& \frac{1}{(1-T)Q^2}\L^2  
&\Longrightarrow {(1-T)Q^2}\sim \L^2 \cr
\mbox{etc.}
}$$ 
The power contribution becomes of order 1 at the edge of phase space,
where the invariant mass of the final-state hadronic system 
is squeezed down to a finite value $\sim \L^2$, so that the process
becomes quasi-elastic and can no longer be treated as hard.
As has already been remarked \cite{AkZak}, it is possible that
such effects, being closely related to the Sudakov form factor,
can be shown to exponentiate and factorize universally.

\section*{Acknowledgements}
GM and BW acknowledge the hospitality of the CERN Theory Division 
while much of this work was performed. We thank G.\ Altarelli,
M.\ Beneke, V.M.\ Braun, S.\ Catani, G.\ Martinelli,
N.G.\ Uraltsev and V.I.\ Zakharov for valuable comments.

\vspace{1.0cm}
Note added: The results of Ref.~\cite{BroKat}, based on
the large-$N_f$ limit, also indicate that the GLS sum
rule should receive only $Q^{-2}$ and $Q^{-4}$ power
corrections, as in Eq.~\re{GLSNP}, and suggest that the
power corrections to the polarized Bjorken sum rule
should be the same.  BW thanks D.J.\ Broadhurst for a
discussion of this point.

\newpage
\bAPP{A}{Drell-Yan cross section}
The real contributions of the moments of the Drell-Yan characteristic 
function are given by
\baeq\eqalign{
& \cFr_N(\eps)
 = 4 \left(\frac{c+\sqrt{c^2-1}}{2\,c}\right)^N \int_0^\infty d\eta 
\cr
&\left\{ 
\frac{\eta\sinh\eta}{\cosh\eta}\cdot\frac{c^N}{(\cosh\eta+c)^{N}}
+ \frac{\eta\sinh\eta}{\cosh\eta}\cdot 
\frac{c^{N+2}}{(\cosh\eta+c)^{N+2}}
 -{\sinh^2\eta}\cdot\frac{c^{N}}{(\cosh\eta+c)^{N+2}} 
\right\}\;.
}\eaeq 
Introducing 
\bminiG{threeint}
I_1(c)&=&4 \int_0^\infty \frac{ d\eta\>\eta\,\sinh\eta}
{\cosh\eta(\cosh\eta+c)}= 
\frac2{c}\left[\,\ln^2\left(c+\sqrt{c^2-1}\right)
+\frac{\pi^2}{4}\,\right]\>;\\
I_2(c)&=& -4\int_0^\infty  \frac{ d\eta\>\sinh^2\eta}{(\cosh\eta+c)^3}
= \frac{2\ln\left(c+\sqrt{c^2-1}\right)}{(c^2-1)^{3/2}}
-\frac{2c}{c^2-1}\>,
\emini
we obtain
\baeq
\cFr_N(\eps)
=\> \left(\frac{c+\sqrt{c^2-1}}{2\,c}\right)^N
\left\{ \left[\, D_{N-1} + D_{N+1}\,\right] I_1(c)
\>+\> \frac{2}{N(N\!+\!1)}\,D_{N-1}\,I_2(c) \right\} .
\eaeq
with 
$$
D_n \equiv \frac{c^{n+1}}{n!}\left(-\frac{d}{dc}\right)^n \>.
$$
For the purpose of extracting the first power correction,
the expressions (\ref{threeint}) may be expanded as
\bmini
 I_1(c) &=& \frac{2\ln^2(2c)}{c} +\frac{\pi^2}{2c}- 
\frac{\ln(2c)}{c^3}  + \cO{c^{-5}} \>; \\
 I_2(c) &=& -\frac2{c} + \frac{2\ln(2c)}{c^3} - \frac2{c^3}
+ \cO{c^{-5}}\>.
\emini
The operator $D_n$ acts as follows:
\beeq
D_n \> \left\{\> c^{-k}\>\right\} 
&=& \frac{\Gamma(k+n)}{\Gamma(k)n!}\>c^{-k+1}\>,\nonumber \\
D_n \, \left\{\> \ln(2c)c^{-k} \>\right\} 
&=& \frac{\Gamma(k+n)}{\Gamma(k)n!}\,c^{-k+1}
\left(\ln(2c)+\psi(k)-\psi(k+n)\right), \nonumber\\
D_n  \left\{\>\ln^2(2c)c^{-k} \>\right\} 
&=& \frac{\Gamma(k+n)}{\Gamma(k)n!}\,c^{-k+1}\left( 
\psi'(k+n)-\psi'(k)+
\left[\,\ln(2c)+\psi(k)-\psi(k+n)\,\right]^2\right).\nonumber
\eeeq
Applying the above results, one arrives at, to first order in $\eps$,
\baeq\eqalign{
\cFr_N &= M_N^{(0)} + \eps\cdot M_N^{(1)} + \ldots \>;\cr
M_N^{(0)} &=  (\ln\eps+2S_1)^2+\pi^2 + 2(\ln\eps+2S_1)
\left[\,\frac1N+\frac1{N+1}\,\right] -4S_2 \>, \cr \cr
M_N^{(1)} &=  
-{N}[(\ln \eps+2S_1)^2+\pi^2] +2(\ln\eps+2S_1)
\left[\,N^2+3N-3+\frac1{N+1}\,\right] \cr
& -6N(N+1) +4NS_2 +16-\frac8{N+1}\>.
}\eaeq
Taking into account the virtual correction, one obtains
\baeq\label{cFDYN}\eqalign{
\cF^{DY}_N(\eps)= 
&= \left(4S_1-3+{\frac {2}{N}}+{\frac {2}{N+1}}\right)\ln\eps \cr
& +{\frac {4\,{\pi }^{2}}{3}}
+4(S_1^2 -S_2)
+\left ({\frac {4}{N+1}}+{\frac {4}{N}}\right )S_1 -\frac72 \cr
& +\eps\left\{ -(N+2)(\ln \eps+2S_1)^2
+ 2(\ln \eps+2S_1)\left[\,N^2+3N-3+\frac1{N+1} +4S_1\,\right] \right. 
\cr
& \left. -6N(N+1) 
+4\left (NS_2   -2 S_1^2\right )
-\left (N-\frac23\right ){\pi }^{2}
 +12 -{\frac {8}{N+1}}\right\}.
}\eaeq
Constructing 
$$
\cK_N(\eps)=\cF^{DY}_N(\eps)-2\cF^{DIS}_N(\eps)\,,
$$
one derives the following expression for the characteristic function
of the $K$-factor in moment space:
$$
\cK_N(\eps)=\hat c_N+\eps(\hat f_N \ln^2\eps +
 \hat g_N \ln\eps +\hat h_N)
$$
where
\baeq\label{cKNeq}\eqalign{
\hat c_N &= 2(S_1^2+ S_2) +\left(-3+\frac2N +\frac2{N+1}\right)S_1
+\frac{4\pi^2}{3}+1-\frac4N-\frac6{N+1}+\frac2{N^2}+\frac2{(N+1)^2} 
\cr
\hat f_N &= -(N+2) \cr
\hat g_N &=2N(N+2)-4(N-2)S_1-22+2\,\frac{19N^2+30N+8}{N(N+1)(N+2)} \cr
\hat h_N &=-4(N-1)S_1^2 + \left(4N^2+10N-28 + 
8\frac{5N^2+8N+2}{N(N+1)(N+2)}
\right)S_1 +4(N-3)S_2 \cr
& -6N^2+4N+18 -\left(N-\frac23\right)\pi^2
-2\,\frac{9N^5+61N^4+122N^3+104N^2+48N+16}{N^2(N+1)^2(N+2)^2} \;.
}\eaeq
Its logarithmic derivative may be written in the following form,
which is suitable for studying the large $N$ regime:
$$
\dot\cK_N(\eps)=\eps\left[\check f_N (\ln\eps+2S_1)^2+
\check g_N (\ln\eps+2S_1)+\check h_N \right]
$$
where
\baeq\eqalign{
\check f_N &= N+2 \cr
\check g_N &= -2\left(N(N+1) +8S_1 -13 
+\frac{19N^2+30N+8}{N(N+1)(N+2)}\right) \cr
\check h_N &= -2\left(N+16-\frac4{N}-\frac2{N+1}-\frac{12}{N+2} 
\right)S_1 
+ 4(N-1)^2 -4(N-3)S_2 +20S_1^2 \cr
& +\left(N-\frac23\right)\pi^2 
-4\frac{5N^5+13N^4+7N^3-10N^2-16N-8}{N^2(N+1)^2(N+2)^2}\;.
}\eaeq
\eAPP

\newpage
\par \vskip 1ex
\noindent{\large\bf References}
\begin{enumerate}
\item\label{hadro}
       B.R.\ Webber, \pl{339}{148}{94};
       see also {\em Proc.\ Summer School on Hadronic Aspects
       of Collider Physics, Zuoz, Switzerland, August 1994}, 
       ed.\ M.P.\ Locher \\ (PSI, Villigen, 1994).
\item\label{DW}
       Yu.L.\ Dokshitzer and B.R.\ Webber, \pl{352}{451}{95}.
\item\label{MW}
       A.V.\ Manohar and M.B.\ Wise, \pl{344}{407}{95}.
\item\label{AkZak}
       R.\ Akhoury and V.I.\ Zakharov, \pl{357}{646}{95};\\
       preprint UM-TH-95-19 [hep-ph/9507253].
\item\label{CS}
       H.\ Contopanagos and G.\ Sterman, \np{419}{77}{94}.
\item\label{KS}
       G.P.\ Korchemsky and G.\ Sterman, \np{437}{415}{95}.
\item\label{BB}
       M.\ Beneke and V.M.\ Braun, \np{454}{253}{95}.
\item\label{NW}
       P.\ Nason and B.R.\ Webber, \np{421}{473}{94}.
\item\label{BNP}
   E. Braaten, S. Narison and A.Pich, \np{373}{58}{92};\\
   A. Pich, {\em Nucl.\ Phys.\ (Proc.\ Suppl.)}~\underline{39B} (1995) 
326.
\item\label{ANR}
   G. Altarelli, P. Nason and G. Ridolfi, \zp{68}{257}{95}.
\item\label{GA}
   G. Altarelli, preprint CERN-TH-95/196.
\item\label{BBB}
   P.\ Ball, M.\ Beneke and V.M.\ Braun, \np{452}{563}{95}.
\item\label{Neu}
   M.\ Neubert, preprint CERN-TH/95-112 [hep-ph/9509432].
\item\label{OPE}
    K. Wilson, \pr{179}{1499}{69}.
\item\label{SVZ}
    M.A. Shifman, A.I. Vainstein and V.I. Zakharov, 
    \np{147}{385, 448, 519}{79}.
\item\label{renormalons}
       For recent reviews and classical references see \\
       V.I. Zakharov, \np{385}{452}{92} and \\
       A.H.\ Mueller, in {\em QCD 20 Years Later}, vol.~1
       (World Scientific, Singapore, 1993).
\item\label{BBZ}
  M. Beneke, V.M. Braun and V.I. Zakharov, \prl{73}{3058}{94}.
\item\label{DS} 
   Yu.L. Dokshitzer and D.V. Shirkov, \zp{67}{449}{95}.
\item\label{DDT}
   Yu.L.\ Dokshitzer, D.I.\ Dyakonov and S.I.\ Troyan,
   \prep{58}{270}{80}.
\item\label{Abel}
  D.J. Broadhurst and Grozin, \pr{52}{4082}{95};\\
  M. Beneke and V.M. Braun, \pl{348}{513}{95}.
\item\label{GL72}
   V.N. Gribov and L.N. Lipatov, \spj{15}{438, 675}{72}.
\item\label{ABCMV}
   D.\ Amati, A.\ Bassetto, M.\ Ciafaloni, G.\ Marchesini
   and G.\ Veneziano,\\ \np{173}{429}{80}.
\item\label{2loop}
  W. Furmanski and R. Petronzio, \pl{97}{437}{80}, \zp{11}{293}{82};
  J.Kalinowski, K. Konishi, P.N. Scharbach and T.R. Taylor, 
  \np{181}{253}{81};
  E.G.Floratos, C. Kunnas and R. Lacaze, \np{192}{417}{81}.
\item\label{CMW}
       S. Catani, G. Marchesini and B.R. Webber, \np{349}{635}{91}.
\item\label{BLM}
       S.J. Brodsky, G.P. Lepage and P.B. Mackenzie, \pr{28}{228}{83}.
\item\label{DKT}
       Yu.L.\ Dokshitzer, in {\em Proc.\ XXVII Recontre de Moriond:
       Perturbative QCD and Hadronic Interactions, March, 1992},
       ed.\ J.\ Tran Than Van (Editions Frontieres, 1992);
       Yu.L.\ Dokshitzer, in
       {\em Proc.\ Int.\ School of Subnuclear Physics, Erice, 1993};
       Yu.L.\ Dokshitzer, V.A.\ Khoze and S.I.\ Troyan, Lund preprints
       LU TP 92--10, 94--23 
       (to appear in Phys.Rev.D, January, 1996).
\item\label{NS}
       P.\ Nason and M.H.\ Seymour, \np{454}{291}{95}.
\item\label{ITEP}
       {\it Vacuum Structure and QCD Sum Rules: Reprints}, 
       ed.\ M.A.\ Shifman \\  
       (North-Holland, 1992:
       Current Physics, Sources and Comments, v.\ 10).
\item\label{DU}
 Yu.L.\ Dokshitzer and N.G. Uraltsev, preprint CERN-TH/95-328
[hep-ph/9512407].
\item\label{DSU}
D. Dikeman, M. Shifman and N.G. Uraltsev,  \ijmp{A11}{571}{96}.
\item\label{LPHD} 
       Yu.L.\ Dokshitzer, V.A.\ Khoze, A.H.\ Mueller and S.I.\ Troyan,
       {\em Basics of Perturbative QCD}, ed. J.\ Tran Thanh Van, 
       (Editions Fronti{\`e}res, 1991).
\item\label{AEM}
       G. Altarelli, R.K. Ellis and G. Martinelli, 
\np{143}{521}{78};\\
       J. Kubar-Andr\'e and F.E. Paige, \pr{19}{221}{79}.
\item\label{VM}
       M.\ Virchaux and A.\ Milstajn, \pl{274}{221}{92}.
\item\label{MRSA}
       A.D. Martin, R.G. Roberts and W.J. Stirling, \pr{50}{6734}{94}.
\item\label{BK}
     V.M.\ Braun and A.V.\ Kolesnichenko, \np{283}{723}{87};
\item\label{harris}
     D.A.\ Harris, CCFR-NuTeV Collaboration, FERMILAB-CONF-95-144,
     to appear in {\em Proc.\ 30th Rencontres de Moriond, Meribel les
     Allues, France, March 1995}. 
\item\label{ALEPHfrag}
       ALEPH Collaboration, D.\ Buskulic et al., \pl{357}{487}{95}.
\item\label{CDW}
     S. Catani, Yu.L. Dokshitzer and B.R. Webber, 
     in preparation.
\item\label{Duralg}
     S. Catani, Yu.L. Dokshitzer, M. Olsson, G. Turnock and B.R. Webber, \\
     \pl{269}{432}{91}.
\item\label{Siggi}
     OPAL Collaboration, M.Z.\ Akrawy et al., \zp{49}{375}{91};\\
     S.\ Bethke, Z.\ Kunszt, D.E.\ Soper and W.J.\ Stirling,
     \np{370}{310}{92}.
\item\label{BroKat}
      D.J.\ Broadhurst and A.L.\ Kataev, \pl{315}{179}{93}.
\end{enumerate}
\end{document}